\documentclass[journal]{IEEEtran}

\usepackage{cite}
\usepackage{graphicx}
\usepackage[cmex10]{amsmath}
\usepackage[justification=centering]{caption}
\usepackage{float}
\usepackage{verbatim}
\usepackage{relsize}
\usepackage{amsfonts}
\usepackage{amsthm}
\usepackage{amssymb}
\usepackage{latexsym}
\usepackage{url}
\usepackage{epstopdf}
\usepackage{enumerate}
\usepackage{mathrsfs}
\usepackage{tabulary, hyperref}
\usepackage{color}
\usepackage{balance}
\usepackage{amsmath}
\usepackage{cases}
\usepackage{caption}

\DeclareMathOperator*{\argmin}{arg\,min}
\newtheorem{theorem}{Theorem}
\newtheorem{lemma}{Lemma}

\newtheorem{proposition}{Proposition}
\newtheorem{remark}{Remark}

\newcommand{\bfc}{{\boldsymbol c}}

\newcommand{\GV}{{\mathrm{GV}}}

\newcommand{\TLC}{{\mathrm{TLC}}}
\newcommand{\LP}{{\mathrm{LP}}}
\newcommand{\I}{{\mathrm{I}}}
\newcommand{\J}{{\mathrm{J}}}
\newcommand{\SP}{{\mathrm{sp}}}

\allowdisplaybreaks
\begin{document}
		
\title{On the Bee-Identification Error Exponent\\ with Absentee Bees}

\author{Anshoo Tandon,~\IEEEmembership{Member,~IEEE}, Vincent Y.\ F.\ Tan,~\IEEEmembership{Senior Member,~IEEE}, \\ and Lav R.\ Varshney,~\IEEEmembership{Senior Member,~IEEE}% <-this % stops a space
	\thanks{This work was supported in part by a Singapore Ministry of Education Tier~2 grant (R-263-000-C83-112).}%
\thanks{A.~Tandon is with the Department of Electrical and Computer Engineering, National University of Singapore, Singapore 117583 (email: anshoo.tandon@gmail.com).}%
\thanks{V.~Y.~F.~Tan is with the Department of Electrical and Computer Engineering, and with the Department of Mathematics, National University of Singapore, Singapore (email: vtan@nus.edu.sg).}%
\thanks{L.~R.~Varshney is with the Coordinated Science Laboratory and the Department of Electrical and Computer Engineering, University of Illinois at Urbana-Champaign, Urbana, IL~61801 USA (email: varshney@illinois.edu)}%
}

\maketitle

\begin{abstract}
The ``bee-identification problem'' was formally defined by Tandon, Tan and Varshney [{\em IEEE Trans. Commun.} (2019) [Online early access]], and the error exponent was studied. This work extends the results for the ``absentee bees'' scenario, where a small fraction of the bees are absent in the beehive image used for identification. For this setting, we present an \emph{exact} characterization of the bee-identification error exponent, and show that independent barcode decoding is optimal, i.e., joint decoding of the bee barcodes does not result in a better error exponent relative to independent decoding of each noisy barcode. This is in contrast to the result \emph{without} absentee bees, where joint barcode decoding results in a significantly higher error exponent than independent barcode decoding. We also define and characterize the ``capacity'' for the bee-identification problem with absentee bees, and prove the strong converse for the same. 
\end{abstract}

\begin{IEEEkeywords}
	Bee-identification problem, absentee bees, noisy channel, error exponent, capacity, strong converse.
\end{IEEEkeywords}

\section{Introduction}

\begin{figure}[t]
	\centering
	\includegraphics[width=0.28\textwidth, angle=90]{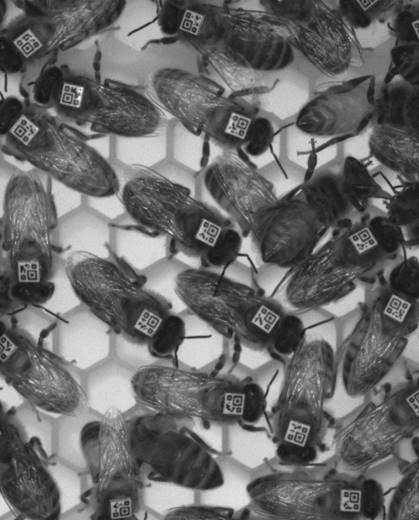}
	\caption{Bees tagged with barcodes\\(photograph provided by T.\ Gernat and G.\ Robinson).}
	\label{Fig:Beehive}
\end{figure}

The problem of bee-identification   with absentee bees can be described as follows. Consider a group of $m$ different bees, in which each bee is tagged with a {\em unique} barcode for identification purposes in order to understand interaction patterns in honeybee social networks~\cite{Tandon19_Bee_TCOM,Gernat18}. Assume a camera takes a picture of the beehive to study the interactions among bees. The beehive image output (see Fig.~\ref{Fig:Beehive}) can be considered as a noisy and unordered set of barcodes. In this work, we consider the ``absentee bees'' scenario, in which some bee barcodes  are missing in the image used to decode the identities of the bees. This scenario can arise, for instance, when some of the bees fly away from the beehive, or when some of the bees (or their barcodes) are occluded from view. Posing as an information-theoretic problem, we quantify the error probability of identifying the bees still present in the finite-resolution beehive image through the corresponding (largest or best) error exponent.

\begin{figure*}[t]
	\centering
	\scalebox{1.0}{ 
		\begin{picture}(380,65)(0,0)
		\put(165,53){Effective Channel}
		\put(-3,35){Codebook {\large$\mathcal{C}$}}
		\thicklines
		\put(15,30){\vector(1,0){50}}
		\put(65,22){\framebox(87,20){Row-Permutation {\large$\pi$}}}
		\put(152,30){\vector(1,0){25}}
		\put(156.5,36){{\large $\mathcal{C}_{\pi}$}}
		\put(177,22){\framebox(63,20){Delete $k$ rows}}
		\put(240,30){\vector(1,0){46}}
		\put(244,37.5){{\large $\mathcal{C}_{\pi_{(m-k)}}$}}
		\put(286,22){\framebox(40,20){BSC($p$)}}
		\put(326,30){\vector(1,0){55}}
		\put(337,37.5){\large$\tilde{\mathcal{C}}_{\pi_{(m-k)}}$}
		\thinlines
		\put(56.5,15){\framebox(274,35){}}
		\end{picture}
	}
	\caption{Effective channel for the bee-identification problem with $k$ absentee bees.}
	\label{Fig:BeeChannelAbsentee}
\end{figure*}
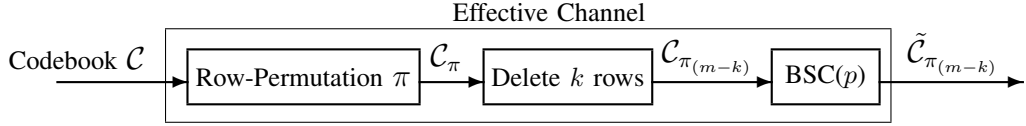

The barcode for each bee is represented as a binary vector of length $n$, and the bee barcodes are collected in a codebook $\mathcal{C}$ comprising $m$ rows and $n$ columns, with each row corresponding to a bee barcode. As shown in Fig.~\ref{Fig:BeeChannelAbsentee}, the channel first permutes the $m$ rows of $\mathcal{C}$ with a random permutation $\pi$ to produce $\mathcal{C}_\pi$, where the $i$-th row of $\mathcal{C}_\pi$ corresponds to the $\pi(i)$-th row of $\mathcal{C}$. Next, the channel deletes $k$ rows of $\mathcal{C}_\pi$, to model the scenario in which $k$ bees, out of a total of $m$ bees, are absent in the beehive image. Without loss of generality, we assume that the channel deletes the last $k$ rows of $\mathcal{C}_\pi$ to 
produce $\mathcal{C}_{\pi_{(m-k)}}$, where $\pi_{(m-k)}$ denotes an injective mapping from $\{1,\ldots,m-k\}$ to $\{1,\ldots,m\}$ and corresponds to the restriction of permutation $\pi$ to only its first $m-k$ entries. Finally, the channel adds noise, modeled as a binary symmetric channel (BSC) with crossover probability $p$  with $0 < p < 0.5$, to produce $\tilde{\mathcal{C}}_{\pi_{(m-k)}}$ at the channel output. We assume the decoder has knowledge of codebook $\mathcal{C}$, and its task is to \emph{recover the channel-induced mapping} $\pi_{(m-k)}$ using the channel output $\tilde{\mathcal{C}}_{\pi_{(m-k)}}$. Note that $\pi_{(m-k)}$ directly ascertains the identity of all $m-k$ bees present in the image.

When $j=\pi(i)$ and the $j$-th row of codebook $\mathcal{C}$ is denoted $\bfc_j = [c_{j,1}~ c_{j,2}~ \cdots~ c_{j,n}]$, then the $i$-th row of $\mathcal{C}_{\pi}$ is equal to $\bfc_j$. For $1 \le i \le m-k$, the $i$-th row of $\tilde{\mathcal{C}}_{\pi_{(m-k)}}$, denoted $\tilde{\bfc}_i$, is a noisy version of $\bfc_j = \bfc_{\pi(i)}$ and we have
\begin{align}
\Pr\{\tilde{\bfc}_{i} \mid \bfc_{\pi(i)} \} &= p^{d_i} (1-p)^{n-{d_i}} ,~~1 \le i \le m-k, \nonumber \\
\Pr\big\{\tilde{\mathcal{C}}_{\pi_{(m-k)}} \,\big|\, \mathcal{C}, \pi_{(m-k)} \big\} &= \prod_{i=1}^{m-k} \Pr\{\tilde{\bfc}_{i} | \bfc_{\pi(i)} \} \nonumber \\
&= \prod_{i=1}^{m-k} p^{d_i} (1-p)^{n-{d_i}},  \label{eq:output_distribution}
\end{align}
where ${d_i} \triangleq \mathrm{d_H}(\tilde{\bfc}_{i},\bfc_{\pi(i)})$ denotes the Hamming distance between vectors $\tilde{\bfc}_i$ and $\bfc_{\pi(i)}$.

We remark that the bee-identification problem formulation has other applications in engineering, such as  package-distribution to recipients from a batch of deliveries with noisy address labels, and identification of warehouse products using wide-area sensors~\cite{Tandon19_Bee_TCOM}. In a related work on identification via permutation recovery~\cite{Shahi18}, the  identification of the respective distributions of a set of observed sequences (in which each sequence is generated i.i.d.\ by a distinct distribution) was analyzed. Other models and applications, in which permutation recovery arises naturally, are discussed in~\cite{Pananjady18}. 

In another related work, the fundamental limit of data storage via unordered DNA molecules was investigated in~\cite{Heckel17}, while the corresponding capacity results were extended to a noisy setting in~\cite{Ilan19ISIT}. Here, we remark that the effective channel in \cite{Ilan19ISIT} is closely related to the bee-identification channel. Data communication over permutation channels with impairments, such as insertions, deletions, and substitutions, was analyzed in~\cite{Mladen18_Multisets}. Note that the effective channel for the bee-identification problem differs from the communication channel in \cite{Mladen18_Multisets} in two aspects: (i) The input to the channel in the bee-identification problem is the entire codebook (see Fig.~\ref{Fig:BeeChannelAbsentee}), instead of just a codeword belonging to the codebook; (ii)~The channel in Fig.~\ref{Fig:BeeChannelAbsentee} only permutes the {\em rows} of the codebook, but does not permute the {\em letters} within a row.

\subsection{Absentee Bee-Identification Error Exponent} \label{Sec:IntroBeeExponent}
The bee-identification problem involves designing a  decoder to  detect the channel-induced mapping $\pi_{(m-k)}$ by using the knowledge of codebook $\mathcal{C}$ and channel output $\tilde{\mathcal{C}}_{\pi_{(m-k)}}$. The decoder corresponds to a function $\phi$ that takes $\tilde{\mathcal{C}}_{\pi_{(m-k)}}$ as an input and produces a map $\nu : \{1,\ldots,m-k\} \to \{1,\ldots,m\}$ where $\nu(i)$ corresponds to the index of the transmitted codeword which produced the received word $\tilde{\bfc}_i$, where $1\le i \le m-k$. The indicator variable for the bee-identification error is defined as
\begin{equation*}
\mathcal{D}\left(\nu,\pi_{(m-k)}\right) \triangleq \begin{cases}
1, ~~~\mbox{if } \nu\neq \pi_{(m-k)}, \\
0, ~~~\mbox{if } \nu=\pi_{(m-k)} .
\end{cases}
\end{equation*}

Let $\Upsilon$ denote the set of all injective maps from $\{1,\ldots,m-k\}$ to $\{1,\ldots,m\}$. For a given codebook $\mathcal{C}$ and decoding function $\phi$, the expected bee-identification error probability over the BSC($p$) is
\begin{equation}
D(\mathcal{C},p,k,\phi) \triangleq \mathbb{E}_{\pi_{(m-k)}} \left[ \mathbb{E}\left[\mathcal{D}\left(\nu,\pi_{(m-k)}\right)\right] \right], \label{eq:Def_D}
\end{equation}
where the inner expectation is over the distribution of $\tilde{\mathcal{C}}_{\pi_{(m-k)}}$ given $\mathcal{C}$ and $\pi_{(m-k)}$ (see \eqref{eq:output_distribution}), and the outer expectation is over the uniform distribution of $\pi_{(m-k)}$ over $\Upsilon$. Note that \eqref{eq:Def_D} can be equivalently expressed as
\begin{equation}
D(\mathcal{C},p,k,\phi) = \mathbb{E}_{\pi_{(m-k)}} \left[\Pr\left\{\nu \neq \pi_{(m-k)}\right\}\right]. \label{eq:Def_D_v2}
\end{equation}

Let $\mathscr{C}(n,m)$ denote the set of all binary codebooks of size $m \times n$, i.e. binary codebooks with $m$ codewords, each having length $n$. Now, for given values of $n$, $m$, and $k$, define the minimum expected bee-identification error probability as
\begin{equation}
\underline{D}(n,m,p,k) \triangleq \min_{\mathcal{C},\phi} D(\mathcal{C},p,k,\phi) , \label{eq:Def_min_D}
\end{equation}
where the minimum is over all codebooks $\mathcal{C} \in \mathscr{C}(n,m)$, and all decoding functions $\phi$. The exponent corresponding to the minimum expected bee-identification error probability is given by $-\tfrac{1}{n} \log \underline{D}(n,m,p,k)$. Note that we   take all logarithms to base $2$, unless stated otherwise.

\subsection{Our Contributions}
We consider the bee-identification problem with a constant fraction of ``absentee bees'', and provide an \emph{exact} characterization of the corresponding error exponent; this is done  via Theorem~\ref{Thm:MainResult}. We show that joint decoding of the bee barcodes does not result in a better error exponent relative to the independent decoding of each noisy barcode. This is in contrast to the result \emph{without} absentee bees~\cite{Tandon19_Bee_TCOM}, where joint barcode decoding results in a significantly higher error exponent than independent barcode decoding.

Secondly, we define and characterize the ``capacity'' (i.e., the supremum of all code rates for which the error probability can be driven to $0$) of the bee-identification problem with absentee bees via Theorem~\ref{Thm:CapBeeProblem}. Further, we prove the {\em strong converse} showing that for rates greater than the capacity, the error probability tends to $1$ as the blocklength (length of barcodes) tends to infinity.

Lastly, via Theorem~\ref{Thm:ErrExp_AlphaToZero}, we show that for low rates, the error exponent when the fraction of absentee bees tends to zero, is strictly lower than the corresponding exponent for the case without absentee bees. This implies a {\em discontinuity} in the error exponent function when the fraction of absentee bees  $\alpha$ tends to $0$, highlighting the dichotomy of its behavior when $\alpha>0$ and  when $\alpha=0$. On the one hand, independent barcode decoding is optimal even when arbitrarily small fraction of bees are absent, whereas on the other hand, joint barcode decoding provides higher exponent when no bees are absent.

\section{Bounds on the Error Probability}
In this section, we present finite-length bounds on the minimum expected bee-identification error probability, $\underline{D}(n,m,p,k)$. The upper bound on $\underline{D}(n,m,p,k)$ is presented in Section~\ref{Sec:ID_UB_ErrProb} using a na\"ive decoding strategy in which each noisy barcode is decoded independently, while the lower bound on $\underline{D}(n,m,p,k)$ is presented in Section~\ref{Sec:JD_LB_ErrProb} using joint maximum likelihood (ML) decoding of barcodes. 

\subsection{Independent decoding based upper bound on $\underline{D}(n,m,p,k)$}
\label{Sec:ID_UB_ErrProb}
We present an upper bound on $\underline{D}(n,m,p,k)$ based on two ideas: (i) independent decoding of each barcode, and (ii) the union bound. Independent barcode decoding is a na\"ive strategy where, for $1 \le i \le m-k$, the decoder picks $\tilde{\bfc}_i$, the $i$-th row of $\tilde{\mathcal{C}}_{\pi_{(m-k)}}$, and then decodes it to $\nu(i) = \argmin_{1 \le j \le m} \mathrm{d_H}(\tilde{\bfc}_{i},\bfc_j)$. If there is more than one codeword at the same minimum Hamming distance from $\tilde{\bfc}_i$, then any one of the corresponding codeword indices is chosen uniformly at random. 

We denote the decoding function $\phi$ corresponding to independent barcode decoding as $\phi_{\I}$. Then, for a given codebook $\mathcal{C}$, it follows from \eqref{eq:Def_D_v2} that 
\begin{align}
D(\mathcal{C},p,k,\phi_{\I}) &= \mathbb{E}_{\pi_{(m-k)}} \left[\Pr\left\{\nu \neq \pi_{(m-k)}\right\}\right], \nonumber \\*
&\le \sum_{i=1}^{m-k} \mathbb{E}_{\pi_{(m-k)}} \left[\Pr\left\{\nu(i) \neq \pi_{(m-k)}(i)\right\}\right] , \label{eq:UnionBound_ID}
\end{align}
where the inequality follows from the union bound and the linearity of the expectation operator. 

For a scenario in which $m$ binary codewords, each having blocklength $n$, are used for transmission of information over a binary symmetric channel  BSC($p$), let $P_{\mathrm{e}}(n,m,p)$ denote the minimum achievable average error probability, where the minimization is over all codebooks $\mathcal{C} \in \mathscr{C}(n,m)$. The following lemma applies \eqref{eq:Def_min_D} and \eqref{eq:UnionBound_ID} to present an upper bound on $\underline{D}(n,m,p,k)$ in terms of $P_{\mathrm{e}}(n,m,p)$.

\begin{lemma} \label{Lem:ErrProb_UB_ID}
	Using independent barcode decoding, the bee-identification error probability $\underline{D}(n,m,p,k)$  can be upper bounded  as follows
	\begin{equation}
	\underline{D}(n,m,p,k) \le \min \left\{1,\, (m-k)\, P_{\mathrm{e}}(n,m,p)\right\}. \label{eq:ID_UB_ErrProb_v2}
	\end{equation}
\end{lemma}
\begin{IEEEproof}
	See Appendix~\ref{App:ErrProb_UB_ID}.
\end{IEEEproof}

%The next subsection presents a lower bound on $\underline{D}(n,m,p,k)$ using joint barcode decoding.

\subsection{Joint decoding based lower bound on $\underline{D}(n,m,p,k)$}
\label{Sec:JD_LB_ErrProb}
Recall $\Upsilon$ denotes the set of all injective maps from $\{1,\ldots,m-k\}$ to $\{1,\ldots,m\}$. With joint ML decoding of barcodes using a given codebook $\mathcal{C}$, the decoding function $\phi$ takes the channel output $\tilde{\mathcal{C}}_{\pi_{(m-k)}}$ as an input, and produces the map
\begin{equation}
\nu = \argmin_{\sigma \in \Upsilon} \mathrm{d_H}(\tilde{\mathcal{C}}_{\pi_{(m-k)}},\mathcal{C}_\sigma), \label{eq:nu_JD_def}
\end{equation}
where $\mathcal{C}_\sigma$ denotes a matrix with $m-k$ rows and $n$ columns whose $i$-th row is equal to the $\sigma(i)$-th row of $\mathcal{C}$, and $\mathrm{d_H}(\tilde{\mathcal{C}}_{\pi_{(m-k)}},\mathcal{C}_\sigma) \triangleq |\{(i,j) : \tilde{\mathcal{C}}_{\pi_{(m-k)}}(i,j) \neq \mathcal{C}_\sigma(i,j), 1\le i \le m-k, 1 \le j \le n\}|$. For this joint ML decoding scheme, we denote the decoding function as $\phi_{\J}$. As $\pi_{(m-k)}$ is uniformly distributed over $\Upsilon$, the joint ML decoder minimizes the error probability~\cite[Thm.~8.1.1]{GallagerBook08}, and from \eqref{eq:Def_min_D} we have
\begin{equation}
\underline{D}(n,m,p,k) = \min_{\mathcal{C} \in \mathscr{C}(n,m)} D(\mathcal{C},p,k,\phi_{\J}). \label{eq:Min_D_v2}
\end{equation}
The following lemma uses \eqref{eq:Min_D_v2} to present a lower bound on $\underline{D}(n,m,p,k)$ in terms of $P_{\mathrm{e}}(n,k+1,p)$.

\begin{lemma} \label{Lem:ErrProb_LB_JD}
Let $0 < \varepsilon < 1/2$, and let $k > 1/\varepsilon$. Then,  the bee-identification error probability $\underline{D}(n,m,p,k)$ using joint ML decoding of barcodes can be lower bounded as follows
	\begin{equation}
	  \underline{D}(n,m,p,k) \! >\!  \frac{1\! -\! 2\varepsilon}{2}\min\big\{1, (m\! -\! k)\varepsilon \, P_{\mathrm{e}}(n, \lfloor k \varepsilon \rfloor ,p) \big\}. \label{eq:JD_ErrProb_LB_v6}
	\end{equation}
Furthermore, the error probability $\underline{D}(n,m,p,k)$ can alternatively be lower bounded as follows
	\begin{align}
	&\underline{D}(n,m,p,k) \nonumber\\
	&\quad > (1-2\varepsilon) \big[1 - \exp\left(-(m-k)\varepsilon \, P_{\mathrm{e}}(n, \lfloor k \varepsilon \rfloor ,p)\right)\big]. \label{eq:JD_ErrProb_LB_v6_new}
	\end{align}
\end{lemma}
\begin{IEEEproof}
	See Appendix~\ref{App:ErrProb_LB_JD}.
\end{IEEEproof}
The lower bound in \eqref{eq:JD_ErrProb_LB_v6} will be used to prove the converse part in Theorem~\ref{Thm:MainResult} on characterizing the error exponent. On the other hand, the lower bound in \eqref{eq:JD_ErrProb_LB_v6_new} helps us to characterize the ``capacity'' of the bee-identification problem in Theorem~\ref{Thm:CapBeeProblem}  and to prove the strong converse for the same problem.

%The following section characterizes the error exponent and capacity for the bee-identification problem with absentee bees, and discusses the optimality of independent barcode decoding.

\section{Bee-Identification Exponent and the Optimality of Independent Decoding}
In this section, we analyze the exponent of the minimum expected bee-identification error probability, $-\tfrac{1}{n} \log \underline{D}(n,m,p,k)$. We first present some notation for the bee-identification exponent. Recall that $P_{\mathrm{e}}(n,m,p)$ denotes the minimum achievable average error probability when $m$ binary codewords, each having blocklength $n$, are used for transmission of information over BSC($p$). For a given $R>0$ and $m = \lceil2^{nR}\rceil$,\footnote{We will remove the ceiling operator subsequently; this does not affect the asymptotic behavior of the error exponent $-\tfrac{1}{n} \log \underline{D}(n,m,p,k)$.} the \emph{reliability function} of the channel BSC($p$) is defined as follows~\cite{GallagerBook68},\footnote{Another popular, though perhaps pessimistic, definition of the reliability function given by Han~\cite{HanBook03} and Csisz\'ar-K\"orner~\cite{CsiszarBook_IIed}, replaces $\limsup$ with $\liminf$ in \eqref{eq:Def_RF}.}
\begin{equation}
E(R, p) \triangleq \limsup_{n \to \infty} -\frac{1}{n} \log P_{\mathrm{e}}(n,2^{nR},p) .\label{eq:Def_RF}
\end{equation}
Let $(R_n)_{n \in \mathbb{N}}$ be a sequence that converges to $R$, and for a fixed $n$ we define 
\begin{equation}
E(n,R_n, p) \triangleq -\frac{1}{n} \log P_{\mathrm{e}}(n,2^{nR_n},p) . \label{eq:ErrExp_Pe_n}
\end{equation}
We will relate $E(n,R_n, p)$ to $E(R,p)$ via Lemma~\ref{Lem:ConvergenceToRF}. However, in order to establish Lemma~\ref{Lem:ConvergenceToRF}, we need the fact that $E(R,p)$ is continuous for $R>0$. We remark that although the continuity of $E(R,p)$ (or $E(R,W)$ for a general discrete memoryless channel $W$), has been discussed in previous literature (see, e.g.,~\cite[Lem.~1]{Haroutunian_69}, \cite[p.~113]{Haroutunian08_FnT}, \cite[Prop.~8]{Merhav15_SWC}), a clear and comprehensive proof appears to be elusive. Note that the scenario where the rate is less than the {\em critical rate} is of particular interest, because it is well known that $E(R,W)$ is continuous (and, in fact, convex) for rates greater than the critical rate~\cite{GallagerBook68}. In Appendix~\ref{App:ContinuityOfRF}, we provide a simple and complete proof of the continuity of the reliability function.

\begin{lemma} \label{Lem:ConvergenceToRF}
Assume that the sequence $(R_n)_{n \in \mathbb{N}}$ converges, and that $R = \lim_{n \to \infty} R_n$. Then we have 
\begin{equation}
\limsup_{n \to \infty} E(n,R_n, p) \, = \, E(R,p). \label{eq:ErrExp_Pe}
\end{equation}
\end{lemma}
\begin{IEEEproof}
	See Appendix~\ref{App:ConvergenceToRF}.
\end{IEEEproof}
Lemma~\ref{Lem:ConvergenceToRF} will be pivotal in establishing the exact bee-identification exponent (via Theorem~\ref{Thm:MainResult}), as well as in characterizing the ``capacity'' of the bee-identification problem (via Theorem~\ref{Thm:CapBeeProblem}).

We will characterize the exact bee-identification error exponent for the following scenario:
\begin{itemize}
	\item For a given $R>0$, the number of bee barcodes $m$ scale exponentially with blocklength $n$ as $m = 2^{nR}$. 
	\item For a given $0 < \alpha < 1$, the number of absentee bees $k$ scale as $k = \lfloor\alpha m\rfloor$,\footnote{We will remove the floor operator subsequently.} where $\alpha$ denotes the fraction of bees missing from the camera image.
\end{itemize}
For this scenario, we define the bee-identification exponent as follows,\footnote{We remark that the result in Theorem~\ref{Thm:MainResult} goes through verbatim if we replace $\limsup$ with $\liminf$ in definitions \eqref{eq:Def_RF} and \eqref{eq:ErrExp_Dmin}.}
\begin{equation}
E_{\underline{D}}(R,p,\alpha) \triangleq \limsup_{n \to \infty} -\frac{1}{n}\log \underline{D}(n,m,p,k). \label{eq:ErrExp_Dmin}
\end{equation}
The following theorem uses Lemmas~\ref{Lem:ErrProb_UB_ID}, \ref{Lem:ErrProb_LB_JD}, and \ref{Lem:ConvergenceToRF}, to establish the main result in this paper.

\begin{theorem} \label{Thm:MainResult}
	For $0 < \alpha < 1$, we have
	\begin{equation}
	E_{\underline{D}}(R,p,\alpha) = |E(R,p) - R|^+, \label{eq:ErrExp_Dmin_Exact}
	\end{equation}
	where $|x|^+ \triangleq \max(0,x)$. Further, this exponent is achieved via independent decoding of each barcode.
\end{theorem}
\begin{IEEEproof}
	See Appendix~\ref{App:MainResult}.
\end{IEEEproof}

The above theorem implies the following remarks.

\begin{remark}
For a given $0<\alpha < 1$, if the number of absentee bees $k$ scales as $\alpha m$, then independent barcode decoding is \emph{optimal}, i.e., independent decoding of barcodes does not lead to any loss in the bee-identification exponent, relative to joint ML decoding of barcodes. This is in contrast to the result in~\cite{Tandon19_Bee_TCOM}, which showed that if no bees are absent, then joint barcode decoding provides significantly better bee-identification exponent relative to independent barcode decoding.
\end{remark}

\begin{remark}
	The lower bound on the bee-identification error probability using joint ML decoding in Lemma~\ref{Lem:ErrProb_LB_JD} was obtained by considering only those events in which just a {\em single} bee is incorrectly identified. The proof of Theorem~\ref{Thm:MainResult} employs Lemma~\ref{Lem:ErrProb_LB_JD}, and implies that these error events dominate the error exponent. 
\end{remark}

\begin{remark}
The bee-identification exponent $E_{\underline{D}}(R,p,\alpha)$ does not depend on the precise value of $0<\alpha<1$.
\end{remark}

\subsection{`Capacity' of the bee-identification problem}
The bee-identification exponent~\eqref{eq:ErrExp_Dmin} is exactly characterized in terms of the reliability function $E(R,p)$ via Theorem~\ref{Thm:MainResult}, when the total number of bees scale as $m = 2^{nR}$ with $R>0$, and the number of absentee bees scale as $k = \alpha m$ with $0 < \alpha < 1$. For this same setting, we now formulate and characterize the ``capacity'' of the bee-identification problem. 

For $0 \le \epsilon < 1$, we say that rate $R$ is \emph{$(\alpha,\epsilon)$-achievable} if $\liminf_{n \to \infty} \underline{D}(n,2^{nR},p,\alpha 2^{nR}) \le \epsilon$, and define the {\em $\epsilon$-capacity} of the bee-identification problem as the supremum of all $(\alpha,\epsilon)$-achievable rates. We denote this $\epsilon$-capacity as\footnote{This is analogous to the {\em optimistic  $\epsilon$-capacity} defined by Chen and Alajaji~\cite[Def.\ 4.10]{ChenAlajaji99}.}
\begin{equation}
 C_{\underline{D}}(p,\alpha,\epsilon) \!\triangleq \!\sup \left\{R \!:\! \liminf_{n \to \infty} \underline{D}(n,2^{nR},p,\alpha 2^{nR}) \!\le\! \epsilon \right\}\!. \!\label{eq:Def_CapBeeProblem}
\end{equation}
The above definition implies that for $R < C_{\underline{D}}(p,\alpha,\epsilon)$, there exists a decoding function $\phi$, and a codebook $\mathcal{C}$ with $2^{nR}$ codewords having blocklength $n$, for which the bee-identification error probability $D(\mathcal{C},p,\alpha  2^{nR}, \phi) < \epsilon$, for infinitely many $n$.

Now, the \emph{Bhattacharyya parameter} for BSC($p$) is~\cite{McElieceOmura77}
\begin{equation}
B_p \triangleq  - \log \sqrt{4p(1-p)}, \label{eq:Def_Bhatta}
\end{equation}
and it is well known that~\cite{McElieceOmura77}
\begin{equation}
\lim_{R \downarrow 0} E(R,p) = \frac{B_p}{2}. \label{eq:Reliability_and_Bhatta}
\end{equation} 
For a given $0 < p < 0.5$, define the function $f(R) \triangleq E(R,p)-R$. From \eqref{eq:Def_Bhatta} and \eqref{eq:Reliability_and_Bhatta}, it follows that $\lim_{R \downarrow 0} f(R) > 0$, while $f(1) = -1$ because $E(R,p)=0$ for $R \ge 1 - H(p)$, where $H(\cdot)$ denotes the binary entropy function. Further, $f(\cdot)$ is continuous because $E(R,p)$ is continuous in $R$ (see Appendix~\ref{App:ContinuityOfRF}). Therefore, it follows from the {\em intermediate value theorem}~\cite{RudinBook} that the equation $f(R) = E(R,p) - R = 0$ has a positive solution, and this solution is unique because $f(R)$ is strictly decreasing in $R$. The following theorem states that the capacity of the bee-identification problem with absentee bees is {\em equal} to the unique solution of the equation $f(R)=0$.

\begin{theorem} \label{Thm:CapBeeProblem}
	For $0 < \alpha < 1$, and every $0 \le \epsilon < 1$, we have
	\begin{equation}
	C_{\underline{D}}(p,\alpha,\epsilon) = R_p^*, \label{eq:CapBeeProblem}
	\end{equation}
	where $R_p^*$ is unique positive solution of the following equation
	\begin{equation}
	E(R,p) = R.	\label{eq:Def_Rp_star}
	\end{equation}
\end{theorem}
\begin{IEEEproof}
	See Appendix~\ref{app:CapBeeProblem}.
\end{IEEEproof}
Theorem~\ref{Thm:CapBeeProblem} and its proof lead to the following remarks.
\begin{remark}
We prove the {\em strong converse property}~\cite{Wolfowitz78} in Appendix~\ref{app:CapBeeProblem}, showing that if $R > R_p^*$, then the error probability $\underline{D}(n,2^{nR},p,\alpha 2^{nR})$ tends to $1$  as $n \to \infty$.
\end{remark}

\begin{remark}
The expression for the $\epsilon$-capacity~\eqref{eq:CapBeeProblem} is independent of the value of $\alpha \in (0,1)$. Note that a similar behavior was observed for the bee-identification exponent~\eqref{eq:ErrExp_Dmin_Exact}.
\end{remark}

\subsection{Computation of the bee-identification error exponent $E_{\underline{D}}(R,p,\alpha)$, and the bee-identification capacity $C_{\underline{D}}(p,\alpha,\epsilon)$}
We have characterized the bee-identification error exponent $E_{\underline{D}}(R,p,\alpha)$, and capacity $C_{\underline{D}}(p,\alpha,\epsilon)$, via Theorem~\ref{Thm:MainResult} and Theorem~\ref{Thm:CapBeeProblem}, respectively. In this subsection, we discuss some computational aspects of $E_{\underline{D}}(R,p,\alpha)$ and $C_{\underline{D}}(p,\alpha,\epsilon)$.

It is clear from \eqref{eq:ErrExp_Dmin_Exact} and \eqref{eq:CapBeeProblem} that explicit calculations of $E_{\underline{D}}(R,p,\alpha)$ and $C_{\underline{D}}(p,\alpha,\epsilon)$ require the knowledge of the reliability function $E(R,p)$ for a range of values of $R$ and $p$. Although the exact value of $E(R,p)$ is not known for all $R$ and $p$, it can be bounded as follows
\begin{equation}
E_{\TLC}(R,p) \le E(R,p) \le E_{\SP}(R,p), \label{eq:BoundReliabilityFunction}
\end{equation}
where $E_{\TLC}(R,p)$ is the exponent using {\em typical linear codes}~\cite{Barg02} that achieves the best known lower bound on $E(R,p)$ at all rates, and $E_{\SP}(R,p)$ is the {\em sphere packing exponent}~\cite{GallagerBook68} for BSC($p$). The exponent $E_{\TLC}(R,p)$ can be explicitly evaluated using the following expression~\cite{Barg02}
\begin{equation}
E_{\TLC}(R,p) \! \triangleq\!  \begin{cases}
\delta_{\GV}(R) B_p  &0 < R \le R_{\mathrm{ex}} \\
R_0 - R  &R_{\mathrm{ex}} \le R \le R_{\mathrm{cr}} \\
D\left(\delta_{\GV}(R) \| p \right) &R_{\mathrm{cr}} \! \le\!  R \! \le\!  1\! -\! H(p)
\end{cases}  ,
\label{eq:ETLC_Def}
\end{equation}
where $\delta_{\GV}(R)$ is the Gilbert-Varshamov (GV) distance~\cite{Barg02} defined as the value of $\delta$ in the interval $[0,0.5]$ with $H(\delta) = 1-R$, and 
\begin{align*}
R_{\mathrm{ex}} &\triangleq 1 - H\left(\frac{\sqrt{4p(1-p)}}{1+\sqrt{4p(1-p)}}\right) ,\\
R_{\mathrm{cr}} &\triangleq 1 - H\left(\frac{\sqrt{p}}{\sqrt{p} + \sqrt{1-p}}\right) ,\\
R_0 &\triangleq 1 - \log\left(1 + \sqrt{4p(1-p)}\right) ,\\
D(x\|y) &\triangleq x\log\frac{x}{y} + (1-x)\log\frac{1-x}{1-y}. 
\end{align*}
The sphere packing exponent is defined as~\cite{Barg02}
\begin{equation}
E_{\SP}(R,p) \triangleq D\left(\delta_{\GV}(R) \| p \right), ~~0 < R \le 1-H(p). \label{eq:Esp_Def}
\end{equation}
From \eqref{eq:BoundReliabilityFunction}, \eqref{eq:ETLC_Def}, and \eqref{eq:Esp_Def}, we observe that $E(R,p) = D\left(\delta_{\GV}(R) \| p \right)$ for $R_{\mathrm{cr}} < R \le 1-H(p)$, and it is well known that $E(R,p)$ is identically zero for $R \ge 1-H(p)$~\cite{GallagerBook68}. The exponent $E_{\TLC}(R,p)$ is equal to the {\em random coding exponent} $E_{\mathrm{r}}(R,p)$~\cite{GallagerBook68} for $R \ge R_{\mathrm{ex}}$, and therefore the random coding exponent is a {\em tight} lower bound on $E(R,p)$ for $R \ge R_{\mathrm{cr}}$. Although it is not so well known, it is also true that for $0.046 < p < 0.5$, the lower bound $E_{\mathrm{r}}(R,p)$ is tight for certain rates {\em strictly} less than the critical rate $R_{\mathrm{cr}}$~\cite[Thm.~17]{Barg05}.

In general, upper and lower bounds on $E_{\underline{D}}(R,p,\alpha)$ and $C_{\underline{D}}(p, \alpha)$ can be obtained via Theorem~\ref{Thm:MainResult} and Theorem~\ref{Thm:CapBeeProblem}, respectively, and employing best known bounds on $E(R,p)$. 

If we define the following minimum distance metrics
\begin{align*}
d^*(n,R) &\triangleq \max_{C \in \mathscr{C}(n,R)} \min_{\substack{\bfc_{i}, \bfc_{j} \in C\\ \bfc_i \neq \bfc_j}} \mathrm{d_H}(\bfc_i,\bfc_j) , \\
\delta^*(n,R) &\triangleq\frac{ d^*(n,R)}{n} , \\
\delta^*(R) &\triangleq \limsup_{n \to \infty} \delta^*(n,R),
\end{align*}
then $E(R,p)$ can also be upper bounded as~\cite{McElieceOmura77}
\begin{equation}
E(R,p) \le \delta^*(R) B_p. \label{eq:RF_UB1}
\end{equation}
The exact value of $\delta^*(R)$ is not known in general, though we know that $\delta^*(0)=0.5$ and $\delta^*(1)=0$~\cite{McElieceOmura77}. The value $\delta^*(R)$ is lower bounded by $\delta_{\GV}(R)$, and can be upper bounded as follows~\cite{McEliece77,Litsyn99}
\begin{equation}
\delta^*(R) \le \delta_{\LP}(R) \triangleq \frac{1}{2} - \sqrt{ \delta_{\GV}(1-R) (1-\delta_{\GV}(1-R)) }. \label{eq:delta_star_UB}
\end{equation}
Combining \eqref{eq:RF_UB1} and \eqref{eq:delta_star_UB}, we observe that $E(R,p)$ can be   upper bounded as follows
\begin{equation}
E(R,p) \le \delta_{\LP}(R) B_p. \label{eq:RF_UB2}
\end{equation}

The following proposition provides an exact and explicit characterization of $E_{\underline{D}}(R,p,\alpha)$ for certain values of $R$ by applying different bounds on $E(R,p)$.
\begin{proposition} \label{prop:ErrExp_AB_Ris0}
	For given $0 < p < 0.5$ and $0 < \alpha < 1$, we have
	\begin{equation}
	\lim_{R \downarrow 0} E_{\underline{D}}(R,p,\alpha) = B_p/2 . \label{eq:ErrExp_AB_Ris0}
	\end{equation}
	Further, we have $E_{\underline{D}}(R,p,\alpha) = 0$ when $R \ge R_{\mathrm{cr}}$.
\end{proposition}
\begin{IEEEproof}
	From \eqref{eq:ErrExp_Dmin_Exact}, \eqref{eq:BoundReliabilityFunction}, and \eqref{eq:ETLC_Def}, we obtain
	\begin{align}
	\lim_{R \downarrow 0} E_{\underline{D}}(R,p,\alpha) &\ge \lim_{R \downarrow 0} \delta_{\GV}(R) B_p  = B_p/2 . \label{eq:ErrExp_AB_Ris0_LB}
	\end{align}
	On the other hand, using \eqref{eq:ErrExp_Dmin_Exact} and \eqref{eq:RF_UB2}, we get
	\begin{align}
	\lim_{R \downarrow 0} E_{\underline{D}}(R,p,\alpha) &\le \lim_{R \downarrow 0} \delta_{\LP}(R) B_p = B_p/2 , \label{eq:ErrExp_AB_Ris0_UB}
	\end{align}
	and the claim in \eqref{eq:ErrExp_AB_Ris0} follows by combining \eqref{eq:ErrExp_AB_Ris0_LB} and \eqref{eq:ErrExp_AB_Ris0_UB}.
	
	Next, we note that $E_{\TLC}(R_{\mathrm{cr}},p) = R_0 - R_{\mathrm{cr}} = D\left(\delta_{\GV}(R_{\mathrm{cr}}) \| p\right) = E_{\SP}(R_{\mathrm{cr}},p)$. Therefore, we have $E(R_{\mathrm{cr}},p) = R_0 - R_{\mathrm{cr}}$, and it follows from \eqref{eq:ErrExp_Dmin_Exact} that
	\begin{align*}
	E_{\underline{D}}(R_{\mathrm{cr}},p,\alpha) &= |R_0 - 2 R_{\mathrm{cr}}|^+ = 0,  \nonumber
	\end{align*}
	where the last equality follows because $R_0 \le 2R_{\mathrm{cr}}$~\cite{Tandon19_Bee_TCOM}. Finally, the fact that $E_{\underline{D}}(R,p,\alpha)$ is non-increasing in $R$ implies that $E_{\underline{D}}(R,p,\alpha) = 0$ for $R \ge R_{\mathrm{cr}}$. We remark that this result, together with Theorem~\ref{Thm:CapBeeProblem}, implies that $C_{\underline{D}}(p, \alpha) < R_{\mathrm{cr}}$.
\end{IEEEproof}

It is known that for small rates, the explicit upper bound on $E(R,p)$ given by \eqref{eq:RF_UB2} is better than the sphere packing bound $E_{\SP}(R,p)$~\cite{McElieceOmura77}. Further improved upper bounds on $E(R,p)$ can be obtained by using the {\em straight line bound}~\cite{Shannon1967_PartII}, which for BSC($p$) implies that for $R_1 < R_2$, the straight line joining $\delta_{\LP}(R_1) B_p$ and $E_{\SP}(R_2,p)$ is an upper bound on $E(R,p)$ for $R \in (R_1,R_2)$~\cite{McElieceOmura77}.

\begin{figure}[t]
	\centering
	\includegraphics[width=0.48\textwidth]{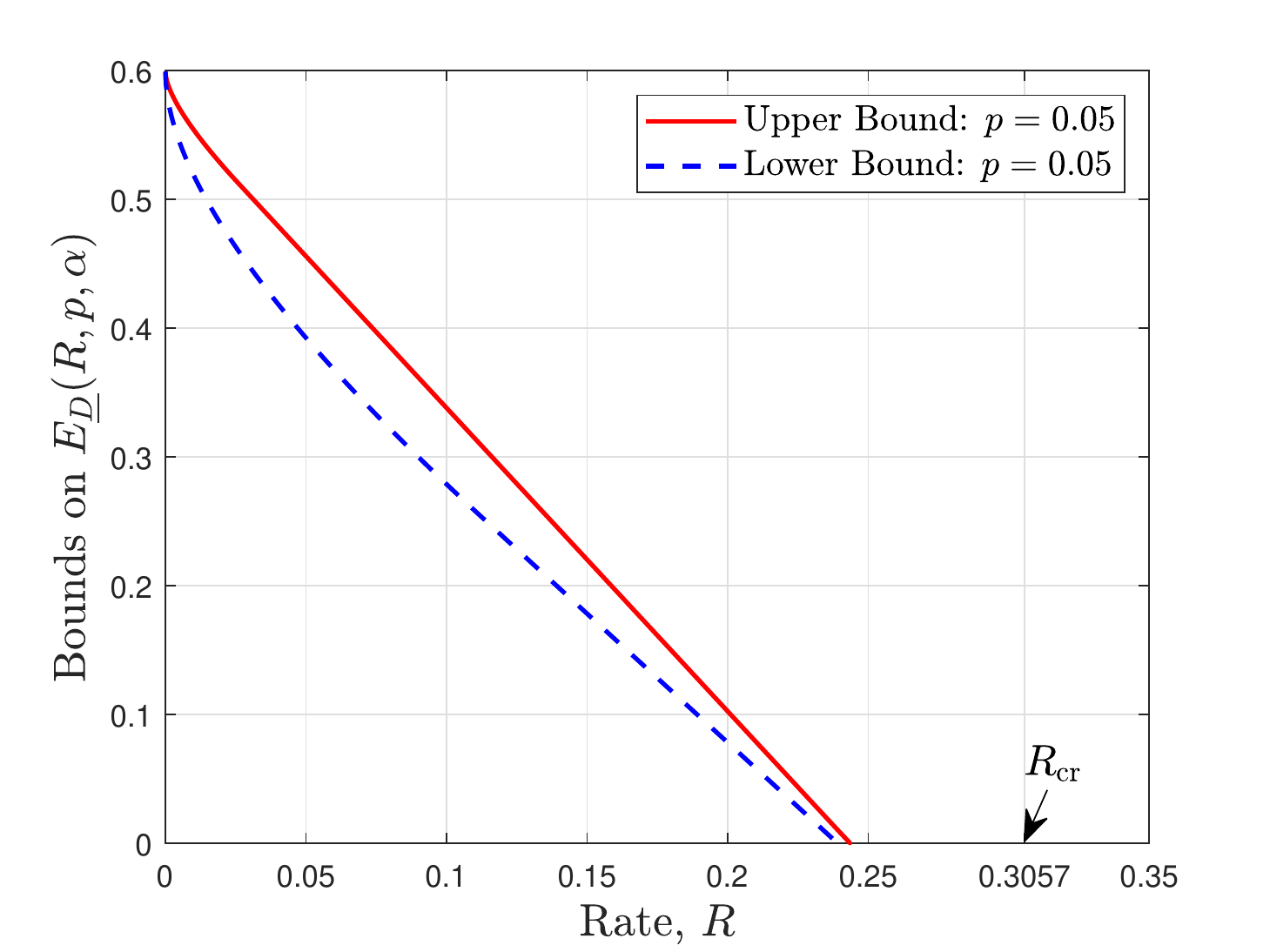}
	\caption{Bounds on $E_{\underline{D}}(R,p,\alpha)$ for $p = 0.05$.}
	\label{Fig:ErrExpBounds_AB}
\end{figure}

Next, we apply the previously discussed bounds on $E(R,p)$ to compute and explicitly bound $E_{\underline{D}}(R,p,\alpha)$ and $C_{\underline{D}}(p, \alpha)$. Fig.~\ref{Fig:ErrExpBounds_AB} plots upper and lower bounds on $E_{\underline{D}}(R,p,\alpha)$ for $p = 0.05$. It is seen from Fig.~\ref{Fig:ErrExpBounds_AB} that the upper and lower bounds coincide as $R$ tends to $0$, and as shown in Prop.~\ref{prop:ErrExp_AB_Ris0}, we have $\lim_{R \downarrow 0} E_{\underline{D}}(R,p,\alpha) = B_p/2 = 0.599$ for $p=0.05$. 

\begin{figure}[t]
	\centering
	\includegraphics[width=0.48\textwidth]{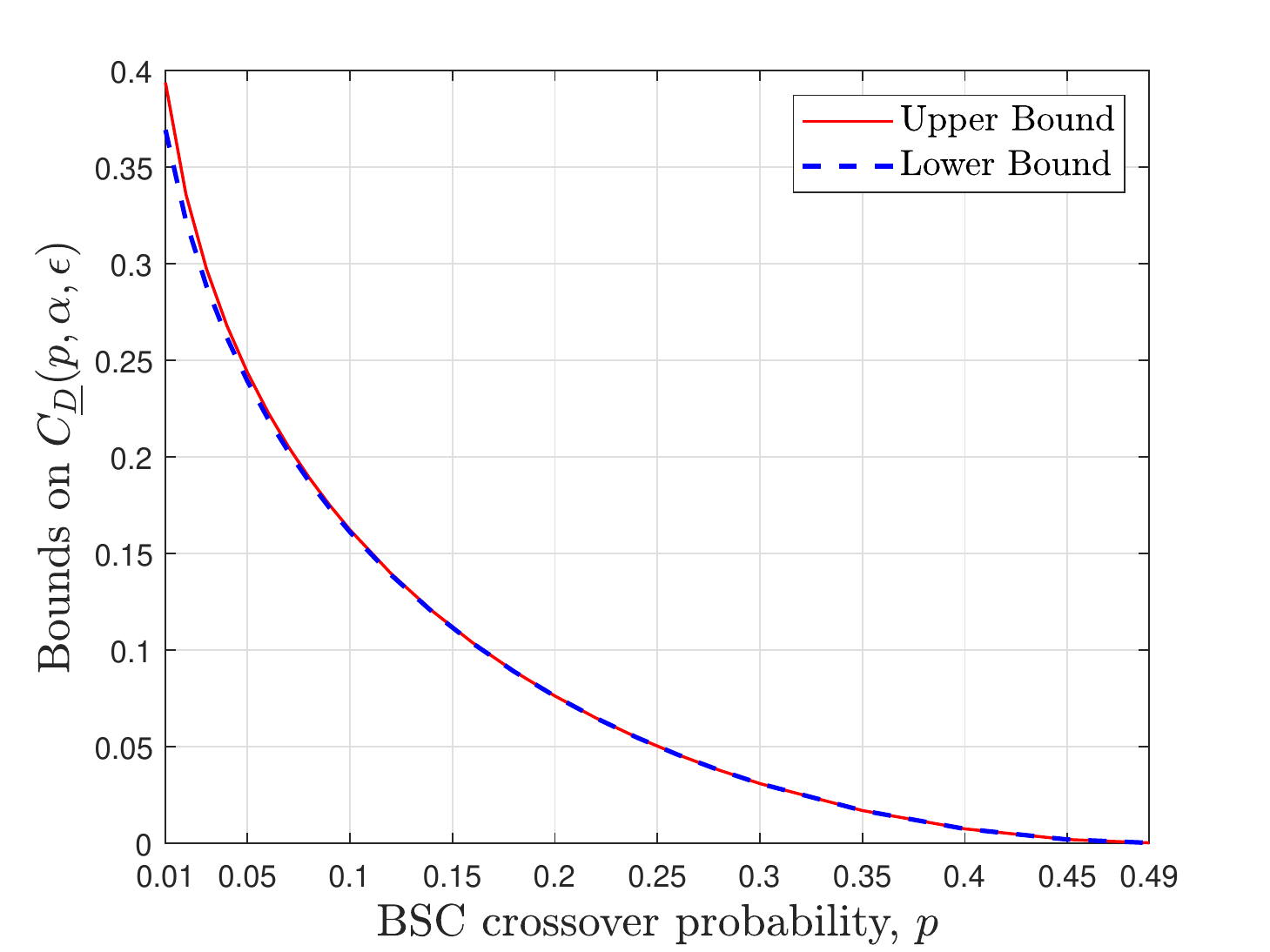}
	\caption{Lower and upper bounds on $C_{\underline{D}}(p,\alpha,\epsilon)$.}
	\label{Fig:CapacityBounds_AB}
\end{figure}

Fig.~\ref{Fig:CapacityBounds_AB} plots lower and upper bounds on the bee-identification capacity $C_{\underline{D}}(p,\alpha,\epsilon)$. As shown in Theorem~\ref{Thm:CapBeeProblem}, when $0 < \alpha < 1$, the capacity is independent of the   value of $\alpha$ (and of $\epsilon$). The numerical results in Fig.~\ref{Fig:CapacityBounds_AB} are obtained using Theorem~\ref{Thm:CapBeeProblem}, and applying the bounds on the reliability function $E(R,p)$, presented in this subsection. It is observed that lower and upper bounds on $C_{\underline{D}}(p,\alpha,\epsilon)$ are relatively close to each other for $p > 0.05$. 

\subsection{Curious case of $\lim_{\alpha \downarrow 0} E_{\underline{D}}(R,p,\alpha)$}
In this section, we analyze the limiting behavior of $E_{\underline{D}}(R,p,\alpha)$ in the setting where $\alpha \downarrow 0$. We will let $E_{\underline{D}}(R,p)$ be the exponent for the {\em no absentee bee} scenario (with $k=0$), and will compare $E_{\underline{D}}(R,p)$ to $\lim_{\alpha \downarrow 0} E_{\underline{D}}(R,p,\alpha)$.

Now, the exponent $E_{\underline{D}}(R,p)$ was studied in detail in~\cite{Tandon19_Bee_TCOM} where upper and lower bounds for the same were derived,\footnote{The exponent $E_{\underline{D}}(R,p)$ was defined in~\cite[Eq.~(5)]{Tandon19_Bee_TCOM} for $m=2^{nR}$ as $\liminf_{n \to \infty} \tfrac{1}{n}\log\left(1/\underline{D}(n,R,p)\right)$. However, the bounds on $E_{\underline{D}}(R,p)$ presented in~\cite{Tandon19_Bee_TCOM} continue to hold if $\liminf$ is replaced by $\limsup$ in the definition of $E_{\underline{D}}(R,p)$.} and it was shown that for $0 < R < 0.5 R_{\mathrm{ex}}$, we have~\cite{Tandon19_Bee_TCOM}
\begin{equation}
2 \delta_{\GV}(2R) B_p  \le E_{\underline{D}}(R,p) \le 2 \delta_{\LP}(R) B_p + R. \label{eq:NAB_ErrExpBounds}
\end{equation}

The next theorem shows that $\lim_{\alpha \downarrow 0} E_{\underline{D}}(R,p,\alpha)$ is {\em strictly less than} $E_{\underline{D}}(R,p,\alpha)$ at low rates.
\begin{theorem} \label{Thm:ErrExp_AlphaToZero}
	For $0 < R < \min\left\{0.169,\, R_{\mathrm{ex}}/2\right\}$, we have the following strict inequality 
	\begin{equation}
	\lim_{\alpha \downarrow 0} E_{\underline{D}}(R,p,\alpha) < E_{\underline{D}}(R,p) .
	\end{equation}
\end{theorem}
\begin{IEEEproof}
	See Appendix~\ref{App:ErrExp_AlphaToZero}.
\end{IEEEproof}

The above result highlights that the limiting behavior for the absentee bee scenario, with $\alpha \downarrow 0$, is quite distinct from the scenario where all bees are present. Independent decoding of bee barcodes is optimal for the absentee bee scenario, even when arbitrarily small fraction of bees are absent. On the other hand, for the scenario where all bees are present, strictly better error exponent, than that obtained by independent decoding, can be achieved via joint ML decoding of barcodes~\cite{Tandon19_Bee_TCOM}.

\subsection{Extension of results to discrete memoryless channels} \label{Subsec:ExtensionToDMC}
In the preceding discussion, we characterized the error exponent and capacity for the bee-identification problem with absentee bees under BSC($p$) noise. This characterization can be readily extended to more general discrete memoryless channels (DMCs). 

Consider a DMC with input alphabet $\mathcal{X}$, output alphabet $\mathcal{Y}$, and channel transition matrix $W$. Then, the reliability function of the DMC, denoted $E(R,W)$, is defined as $
E(R,W) \triangleq \limsup_{n \to \infty} -\tfrac{1}{n} \log P_{\mathrm{e}}(n,2^{nR},W)$, where $P_{\mathrm{e}}(n,2^{nR},W)$ denotes the minimum error probability over all  length-$n$ block codes with $2^{nR}$ codewords. Analogous to~\eqref{eq:Def_min_D}, we may let $\underline{D}(n,m,W,k)$ be the minimum expected bee-identification error probability over the DMC $W$, and define the corresponding bee-identification exponent as
\begin{equation*}
E_{\underline{D}}(R,W,\alpha) \triangleq \limsup_{n \to \infty} -\frac{1}{n}\log \underline{D}(n,2^{nR},W,\alpha 2^{nR}).
\end{equation*}

Let us restrict our attention to DMCs with the property that  there exists an output symbol $y \in \mathcal{Y}$ that is reachable from all symbols in $\mathcal{X}$. This property ensures that $E(R,W)$ is continuous for each $R>0$ (see Appendix~\ref{App:ContinuityOfRF}). The continuity of $E(R,W)$ can be applied to obtain the following result, equivalent to Theorem~\ref{Thm:MainResult} for BSC($p$), as
\begin{equation*}
E_{\underline{D}}(R,W,\alpha) = |E(R,W) - R|^+ .
\end{equation*}
If, analogous to \eqref{eq:Def_CapBeeProblem}, we define the bee-identification $\epsilon$-capacity over DMC $W$ as
\begin{equation*}
C_{\underline{D}}(W,\alpha,\epsilon) \triangleq \sup \left\{R : \liminf_{n \to \infty} \underline{D}(n,2^{nR},W,\alpha 2^{nR}) \le \epsilon \right\},
\end{equation*}
then, for $0 \le \epsilon < 1$, the $\epsilon$-capacity $C_{\underline{D}}(W,\alpha,\epsilon)$ is equal to the unique positive solution of the equation $E(R,W) = R$. Note that this result also uses the continuity of $E(R,W)$, and extends the result in Theorem~\ref{Thm:CapBeeProblem} to DMCs.

\section{Reflections}
This work extended the characterization of the bee-identification error exponent to the ``absentee bees'' scenario, where a fraction of the bees are absent in the beehive image. For this scenario, we presented the \emph{exact} characterization of the bee-identification error exponent in terms of the well known \emph{reliability function}~\cite{GallagerBook68}.

The derivation of the bee-identification exponent led to three interesting observations. The first observation is that when the number of absentee bees $k$ scales as $k = \alpha m$, where $\alpha$ lies in the interval $(0,1)$ and is fixed, and the number of bees $m$ scales exponentially with blocklength, then independent barcode decoding is \emph{optimal}, i.e., joint decoding of the bee barcodes does not result in any better error exponent relative to the independent decoding of each noisy barcode. This result is in contrast to the result \emph{without} absentee bees~\cite{Tandon19_Bee_TCOM}, where joint barcode decoding results in significantly higher error exponent compared to independent barcode decoding. The second interesting observation is that when $k = \alpha m$, the bee-identification exponent is dominated by the events where a {\em single} bee in the beehive image is incorrectly identified as one of the absentee bees, while the other bee barcodes are correctly decoded. The third observation is that for $k=\alpha m$, the bee-identification exponent does not depend on the actual value of $\alpha$ when $0 < \alpha < 1$.

We also characterized the {\em exact} ``capacity'' for the bee-identification problem with absentee bees, and proved the {\em strong converse}. Further, we showed that for low rates, the error exponent for the case where $\alpha \downarrow 0$ is strictly lower than the corresponding error exponent for the case {\em without} absentee bees, thereby highlighting a {\em discontinuity} in the error exponent function at $\alpha = 0$.

The extension of the results presented in this work to general DMCs was briefly discussed in Section~\ref{Subsec:ExtensionToDMC}. Future work includes exploring the error exponent for the scenario where $\alpha$, the fraction of absentee bees, also varies with blocklength $n$, and second-order or finite-length analysis, i.e., the scaling of  the code rate  when $0\le\epsilon<1$ and $n$ is finite. 

\appendices

\section{Proof of Lemma~\ref{Lem:ErrProb_UB_ID}} \label{App:ErrProb_UB_ID}
\begin{IEEEproof}
	For a given codebook $\mathcal{C}$, and given $\pi_{(m-k)}$ and $1 \le i \le m-k$, the probability $\Pr\left\{\nu(i) \neq \pi_{(m-k)}(i)\right\}$ in \eqref{eq:UnionBound_ID} is the probability that the codeword $\bfc_{\pi(i)}$ transmitted over BSC($p$) is incorrectly decoded at the receiver. As $\pi_{(m-k)}$ is uniformly distributed over $\Upsilon$, we have for $1 \le i \le m-k$,
	\begin{equation}
	\min_{\mathcal{C} \in \mathscr{C}(n,m)} \mathbb{E}_{\pi_{(m-k)}} \left[\Pr\left\{\nu(i) \neq \pi_{(m-k)}(i)\right\}\right] = P_{\mathrm{e}}(n,m,p). \label{eq:ID_MinErrPerCodeword}
	\end{equation}
	
	Now, from the definition of $\underline{D}(n,m,p,k)$ in  \eqref{eq:Def_min_D}, we get
	\begin{align}
	\underline{D}(n,m,p,k) &\le \min_{\mathcal{C} \in \mathscr{C}(n,m)} D(\mathcal{C},p,k,\phi_{\I}) \nonumber \\
	&\le (m-k)\, P_{\mathrm{e}}(n,m,p), \label{eq:ID_UB_ErrProb}
	\end{align}
	where \eqref{eq:ID_UB_ErrProb} follows by combining \eqref{eq:UnionBound_ID} and \eqref{eq:ID_MinErrPerCodeword}. Finally, the lemma is proved by using \eqref{eq:ID_UB_ErrProb}, and noting that the bee-identification error probability $\underline{D}(n,m,p,k)$ is trivially upper bounded by 1.
\end{IEEEproof}

\section{Proof of Lemma~\ref{Lem:ErrProb_LB_JD}} \label{App:ErrProb_LB_JD}
\begin{IEEEproof}
Let $\mathcal{I}$ denote the image of $\pi_{(m-k)}$, i.e.,
\begin{equation}
\mathcal{I} \triangleq \{j : j = \pi_{(m-k)}(i), 1 \le i \le m-k\}. \label{eq:Def_mathcalI}
\end{equation}
Let the complement of $\mathcal{I}$ be denoted $\overline{\mathcal{I}}$, i.e.,
\begin{equation}
\overline{\mathcal{I}} \triangleq \{1,2,\ldots,m\} \setminus \mathcal{I}. \label{eq:Def_I_complement}
\end{equation}
For a given $\ell \in \{1,\ldots,m-k\}$ and $\jmath \in \overline{\mathcal{I}}$, define the map $\pi_{(m-k)}^{(\ell \to \jmath)} : \{1,\ldots,m-k\} \to \{1,\dots,m\}$ as follows
\begin{equation*}
\pi_{(m-k)}^{(\ell \to \jmath)}(i) \triangleq 
\begin{cases}
\jmath, &i = \ell \\
\pi_{(m-k)}(i), &i \neq \ell .
\end{cases}
\end{equation*}
Thus, the map $\pi_{(m-k)}^{(\ell \to \jmath)}$ differs with $\pi_{(m-k)}$ only at $i = \ell$, and we have $\pi_{(m-k)}^{(\ell \to \jmath)} \in \Upsilon$. Now, define the event 
\begin{equation*}
\left\{\pi_{(m-k)} \to \pi_{(m-k)}^{(\ell \to \jmath)}\right\} \triangleq \left\{\mathrm{d_H}\left(\tilde{\bfc}_\ell, \bfc_{\jmath}\right) \le \mathrm{d_H}\left(\tilde{\bfc}_\ell, \bfc_{\pi_{(m-k)}(\ell)}\right) \right\},
\end{equation*}
where the error event $\left\{\pi_{(m-k)} \to \pi_{(m-k)}^{(\ell \to \jmath)}\right\}$ implies that only a single bee is decoded incorrectly. Further, if we define the event $\mathscr{E}_{\pi_{(m-k)}}^{(\ell)}$ as
\begin{equation}
\mathscr{E}_{\pi_{(m-k)}}^{(\ell)} \triangleq \bigcup_{\jmath \in \overline{\mathcal{I}}} \left\{\pi_{(m-k)} \to \pi_{(m-k)}^{(\ell \to \jmath)}\right\}, \label{eq:Def_ErrorEvent_l}
\end{equation} 
then, for $\ell \in \{1,\ldots,m-k\}$, we have
\begin{align}
\Pr&\left\{ \mathscr{E}_{\pi_{(m-k)}}^{(\ell)} \right\} \nonumber \\*
&= \Pr\left\{\bigcup_{\jmath \in \overline{\mathcal{I}}} \left\{\mathrm{d_H}\left(\tilde{\bfc}_\ell, \bfc_{\jmath}\right) \le \mathrm{d_H}\left(\tilde{\bfc}_\ell, \bfc_{\pi_{(m-k)}(\ell)}\right) \right\} \right\}. \label{eq:pi_ell_eq}
\end{align}
The output of the ML decoding function $\phi_{\J}$ is given by \eqref{eq:nu_JD_def}, and hence the bee-identification error probability is lower bounded as follows
\begin{equation}
\Pr\left\{\nu \neq \pi_{(m-k)}\right\} \ge \Pr\left\{\bigcup_{1\le \ell \le m-k} \mathscr{E}_{\pi_{(m-k)}}^{(\ell)} \right\}. \label{eq:JD_ErrProb_LB_v2}
\end{equation}
For a given codebook $\mathcal{C}$, we observe from \eqref{eq:pi_ell_eq} that the event $\mathscr{E}_{\pi_{(m-k)}}^{(\ell)}$ depends only on the noise in the $\ell$-th received barcode $\tilde{\bfc}_{\ell}$. Thus, the set of events $\mathscr{E}_{\pi_{(m-k)}}^{(\ell)}$, for $\ell \in \{1,\ldots,m-k\}$, are {\em mutually independent}. Therefore, the probability of their union can be lower bounded using Shulman's lower bound as~\cite[Eq.~(30)]{Merhav13}, \cite[p.~109]{ShulmanThesis}
\begin{equation}
\Pr\left\{\bigcup_{1\le \ell \le m-k} \!\!\mathscr{E}_{\pi_{(m-k)}}^{(\ell)} \right\}\! \ge\! \frac{1}{2} \cdot \min\left\{1, \sum_{\ell = 1}^{m-k}\! \Pr\left\{\mathscr{E}_{\pi_{(m-k)}}^{(\ell)}\right\} \right\}. \label{eq:JD_ErrProb_LB_v3}
\end{equation} 
As $|\overline{\mathcal{I}}| = k$, we observe from \eqref{eq:pi_ell_eq} that the error event $\mathscr{E}_{\pi_{(m-k)}}^{(\ell)}$ occurs when the received word $\tilde{\bfc}_\ell$ is incorrectly decoded to one of the $k$ incorrect codewords $\left\{\bfc_{\jmath}\right\}_{\jmath \in \overline{\mathcal{I}}}$, instead of the correct codeword $\bfc_{\pi_{(m-k)}(\ell)}$, and so 
\begin{equation}
\mathbb{E}_{\pi_{(m-k)}} \left[\Pr\left\{ \mathscr{E}_{\pi_{(m-k)}}^{(\ell)} \right\} \right] \ge P_{\mathrm{e}}(n,k+1,p) \ge P_{\mathrm{e}}(n,k,p) ,\label{eq:JD_ErrProb_LB_v1}
\end{equation}
because $P_{\mathrm{e}}(n,k+1,p)$ denotes the average error probability, minimized over all codebooks with only $k+1$ codewords. Note that \eqref{eq:JD_ErrProb_LB_v1} holds for all codebooks $\mathcal{C} \in \mathscr{C}(n,m)$. Now recall that $\Upsilon$ is the set of all injective maps from $\{1,\ldots,m-k\}$ to $\{1,\ldots,m\}$, and that $\pi_{(m-k)}$ is uniformly distributed over $\Upsilon$. Let $0 < \varepsilon < 1/2$ and $k > 1/\varepsilon$, and define the set
\begin{equation*}
\mathcal{A}^{(\ell)} \triangleq \left\{\pi_{(m-k)} \in \Upsilon \, : \,\Pr\{ \mathscr{E}_{\pi_{(m-k)}}^{(\ell)} \! \} \ge P_{\mathrm{e}}(n, \lfloor k \varepsilon \rfloor ,p)  \right\} .
\end{equation*}
A {\em key observation} is that the size $|\mathcal{A}^{(\ell)}|$ can be bounded as
\begin{equation}
|\mathcal{A}^{(\ell)}| > (1-\varepsilon) |\Upsilon|, \label{eq:Size_Aell_LB}
\end{equation}
where the inequality holds for all  $\ell \in \{1,\ldots,m-k\}$, and all codebooks $\mathcal{C} \in \mathscr{C}(n,m)$. This claim can be explained as follows. First, fix the following variables: codebook $\mathcal{C} \in \mathscr{C}(n,m)$, index $\ell \in \{1,\ldots,m-k\}$, and  $\pi_{(m-k)} \in \Upsilon$. Note that fixing $\pi_{(m-k)}$ in turn fixes the sets $\mathcal{I}$ and $\overline{\mathcal{I}}$, defined in~\eqref{eq:Def_mathcalI} and \eqref{eq:Def_I_complement}, respectively. Let $\overline{\mathcal{I}} = \{\jmath_1, \ldots, \jmath_k\}$, and define
\begin{equation*}
q_{\jmath_r}^{(\ell)} = \Pr\left\{ \mathscr{E}_{\pi_{(m-k)}^{(\ell \to \jmath_r)}}^{(\ell)} \right\} , \;\mbox{ for }\, r \in \{1,\ldots,k\} .
\end{equation*}
Now, let $(\tilde{\jmath}_1 \, \tilde{\jmath}_2 \, \ldots \, \tilde{\jmath}_k)$ be a permutation of the indices $(\jmath_1 \, \jmath_2 \, \ldots \, \jmath_k)$ such that we have the following relation
\[
q_{\tilde{\jmath}_1}^{(\ell)} \le q_{\tilde{\jmath}_2}^{(\ell)} \le \cdots \le q_{\tilde{\jmath}_k}^{(\ell)} .
\]
Thus, $q_{\tilde{\jmath}_{\lfloor k \varepsilon \rfloor}}^{(\ell)}$ satisfies the following property
\[
q_{\tilde{\jmath}_{\lfloor k \varepsilon \rfloor}}^{(\ell)} = \max\left\{q_{\tilde{\jmath}_1}^{(\ell)}, q_{\tilde{\jmath}_2}^{(\ell)}, \ldots , q_{\tilde{\jmath}_{\lfloor k \varepsilon \rfloor}}^{(\ell)} \right\},
\]
and hence
\[
q_{\tilde{\jmath}_{\lfloor k \varepsilon \rfloor}}^{(\ell)} \ge P_{\mathrm{e}}(n, \lfloor k \varepsilon \rfloor ,p) .
\]
The above relation is satisfied because each of $q_{\tilde{\jmath}_1}^{(\ell)}, q_{\tilde{\jmath}_2}^{(\ell)}, \ldots , q_{\tilde{\jmath}_{\lfloor k \varepsilon \rfloor}}^{(\ell)}$ is obtained by comparison among codewords, $\bfc_{\pi_{(m-k)}(\ell)}, \bfc_{\jmath_1}, \bfc_{\jmath_2}, \ldots, \bfc_{\jmath_k}$, while $P_{\mathrm{e}}(n, \lfloor k \varepsilon \rfloor ,p)$ is the minimum achievable average error probability using only $\lfloor k \varepsilon \rfloor$ codewords. Therefore, the fraction of entries in the set $\left\{q_{\jmath_1}^{(\ell)}, q_{\jmath_2}^{(\ell)}, \ldots, q_{\jmath_k}^{(\ell)}\right\}$ that satisfy
\[
q_{\jmath_r}^{(\ell)} = \Pr\left\{ \mathscr{E}_{\pi_{(m-k)}^{(\ell \to \jmath_r)}}^{(\ell)} \right\} \ge P_{\mathrm{e}}(n, \lfloor k \varepsilon \rfloor ,p) ,
\]
is {\em at least} $(k - \lfloor k \varepsilon \rfloor + 1)/k  > (1-\varepsilon)$. This technique can be reapplied to other mappings in $\Upsilon$ to obtain \eqref{eq:Size_Aell_LB}.

Next, construct a matrix $\mathcal{M}$ whose rows are indexed by elements of $\Upsilon$, and the columns are indexed by $\ell \in \{1, 2, \ldots, m-k\}$. For a given row of $\mathcal{M}$ indexed by $\pi_{(m-k)} \in \Upsilon$, let the $\ell$-th entry be equal to $\Pr\left\{ \mathscr{E}_{\pi_{(m-k)}}^{(\ell)} \right\}$. Then, from \eqref{eq:Size_Aell_LB} it follows that {\em at least} $1-\varepsilon$ fraction of entries in each column are lower bounded by $P_{\mathrm{e}}(n, \lfloor k \varepsilon \rfloor ,p)$. Thus, the fraction of entries of matrix $\mathcal{M}$ that are lower bounded by $P_{\mathrm{e}}(n, \lfloor k \varepsilon \rfloor ,p)$ is at least $1-\varepsilon$. Now, we call a row of $\mathcal{M}$ to be $\varepsilon$-{\em strong} if the fraction of entries lower bounded by $P_{\mathrm{e}}(n, \lfloor k \varepsilon \rfloor ,p)$ in that row exceed $\varepsilon$. Let $\theta_{\varepsilon}$ denote the fraction of rows of $\mathcal{M}$ that are $\varepsilon$-strong. Then we have
\begin{align}
(1-\theta_{\varepsilon})\varepsilon + \theta_{\varepsilon} &\ge 1 - \varepsilon, \nonumber 
\end{align}
which implies that 
\begin{align}\theta_{\varepsilon} \ge \frac{1-2\varepsilon}{1-\varepsilon} &> 1-2\varepsilon . \label{eq:ThetaVarepsilon_LB}
\end{align}
Now, we define
\begin{align}
\Upsilon_{\varepsilon} \triangleq \bigg\{\pi_{(m-k)} \in \Upsilon \, : &\, \nonumber \\
 \quad\sum_{\ell = 1}^{m-k} \Pr\big\{ \mathscr{E}_{\pi_{(m-k)}}^{(\ell)} \big\} &> (m-k)\varepsilon \, P_{\mathrm{e}}(n, \lfloor k \varepsilon \rfloor ,p)  \bigg\} , \label{eq:Def_UpsilonVarepsilon}
\end{align}
and note that the elements of $\Upsilon_{\varepsilon}$ correspond to the rows of $\mathcal{M}$ whose row-sum is greater than $(m-k)\varepsilon \, P_{\mathrm{e}}(n, \lfloor k \varepsilon \rfloor ,p)$. Now, we observe that if $\pi_{(m-k)}$ corresponds to a row of $\mathcal{M}$ that is $\varepsilon$-strong, then $\pi_{(m-k)} \in \Upsilon_{\varepsilon}$. Therefore, we have
\begin{equation}
|\Upsilon_{\varepsilon}| \ge \theta_{\varepsilon} |\Upsilon| > (1-2\varepsilon) |\Upsilon|, \label{eq:SizeUpsilonVarepsilon_LB}
\end{equation}
where the strict inequality follows from \eqref{eq:ThetaVarepsilon_LB}. Finally, we have
\begin{align}
&\mathbb{E}_{\pi_{(m-k)}} \left[\Pr\left\{\nu \neq \pi_{(m-k)}\right\}\right] \nonumber \\
&~~\overset{\mathrm{(i)}}{=} \frac{1}{|\Upsilon|} \sum_{\pi_{(m-k)} \in \Upsilon} \Pr\left\{\nu \neq \pi_{(m-k)}\right\} , \nonumber \\
&~~\overset{\mathrm{(ii)}}{\ge} \frac{1}{|\Upsilon|} \sum_{\pi_{(m-k)} \in \Upsilon} \frac{1}{2} \cdot \min\left\{1,\, \sum_{\ell = 1}^{m-k} \Pr\left\{\mathscr{E}_{\pi_{(m-k)}}^{(\ell)}\right\} \right\}, \nonumber  \\
&~~\overset{\mathrm{(iii)}}{\ge} \frac{1}{|\Upsilon|} \sum_{\pi_{(m-k)} \in \Upsilon_{\varepsilon}} \frac{1}{2} \cdot \min\left\{1,\, \sum_{\ell = 1}^{m-k} \Pr\left\{\mathscr{E}_{\pi_{(m-k)}}^{(\ell)}\right\} \right\},  \nonumber \\
&~~\overset{\mathrm{(iv)}}{\ge} \frac{1}{|\Upsilon|} \sum_{\pi_{(m-k)} \in \Upsilon_{\varepsilon}} \frac{1}{2} \cdot \min\left\{1,\, (m-k)\varepsilon \, P_{\mathrm{e}}(n, \lfloor k \varepsilon \rfloor ,p) \right\},  \nonumber \\
&~~\overset{\mathrm{(v)}}{>} \frac{1-2\varepsilon}{2} \cdot \min\left\{1,\, (m-k)\varepsilon \, P_{\mathrm{e}}(n, \lfloor k \varepsilon \rfloor ,p) \right\}, \label{eq:ErrPob_AB_LB}
\end{align}
where $\mathrm{(i)}$ follows because $\pi_{(m-k)}$ is uniformly distributed over $\Upsilon$,\, $\mathrm{(ii)}$ follows from combining \eqref{eq:JD_ErrProb_LB_v2} and \eqref{eq:JD_ErrProb_LB_v3}, \, $\mathrm{(iii)}$ follows because we restrict $\pi_{(m-k)}$ to belong to $\Upsilon_{\varepsilon} \subseteq \Upsilon$,\, $\mathrm{(iv)}$ follows using \eqref{eq:Def_UpsilonVarepsilon} as $\sum_{\ell = 1}^{m-k}\Pr\{ \mathscr{E}_{\pi_{(m-k)}}^{(\ell)}\!\} > (m-k)\varepsilon \, P_{\mathrm{e}}(n, \lfloor k \varepsilon \rfloor ,p)$ for every $\pi_{(m-k)} \in \Upsilon_{\varepsilon}$, and $\mathrm{(v)}$ follows from the fact that $|\Upsilon_{\varepsilon}| / |\Upsilon| >  1-2\varepsilon$ via \eqref{eq:SizeUpsilonVarepsilon_LB}. Finally, we obtain \eqref{eq:JD_ErrProb_LB_v6} by combining \eqref{eq:Def_D_v2}, \eqref{eq:Min_D_v2}, with the fact that \eqref{eq:ErrPob_AB_LB} holds for all codebooks $\mathcal{C} \in \mathscr{C}(n,m)$.

We now proceed to prove the alternative (and stronger) bound in~\eqref{eq:JD_ErrProb_LB_v6_new}. Choose a codebook $\mathcal{C} \in \mathscr{C}(n,m)$ and a mapping $\pi_{(m-k)} \in \Upsilon$. For $\ell \in \{1,\ldots,m-k\}$, let the error event $\mathscr{E}_{\pi_{(m-k)}}^{(\ell)}$ be given by~\eqref{eq:Def_ErrorEvent_l}. Applying \eqref{eq:JD_ErrProb_LB_v2}, we get
\begin{align}
&\Pr\left\{\nu \neq \pi_{(m-k)}\right\} \nonumber\\
&\ge \Pr\left\{\bigcup_{1\le \ell \le m-k} \mathscr{E}_{\pi_{(m-k)}}^{(\ell)} \right\} \nonumber \\
&\overset{\mathrm{(vi)}}{=} 1 - \prod_{\ell = 1}^{m-k}\left(1 - \Pr\left\{ \mathscr{E}_{\pi_{(m-k)}}^{(\ell)} \right\} \right), \nonumber \\
&= 1 - \exp\left(\sum_{\ell = 1}^{m-k} \ln \left(1-\Pr\left\{ \mathscr{E}_{\pi_{(m-k)}}^{(\ell)} \right\}\right)\right), \nonumber \\
&\overset{\mathrm{(vii)}}{\ge} 1 - \exp\left(- \sum_{\ell = 1}^{m-k} \Pr\left\{ \mathscr{E}_{\pi_{(m-k)}}^{(\ell)} \right\}\right), \label{eq:UnionOfEvents_LB}
\end{align}
where $\mathrm{(vi)}$ follows because the events $\mathscr{E}_{\pi_{(m-k)}}^{(\ell)}$, for $\ell \in \{1,\ldots,m-k\}$, are mutually independent, and $\mathrm{(vii)}$ follows because $\ln(1-x) \le -x$ for $x \in [0,1)$. Now, we have
\begin{align}
&\mathbb{E}_{\pi_{(m-k)}} \left[\Pr\left\{\nu \neq \pi_{(m-k)}\right\}\right] \nonumber \\
&~~= \frac{1}{|\Upsilon|} \sum_{\pi_{(m-k)} \in \Upsilon} \Pr\left\{\nu \neq \pi_{(m-k)}\right\} , \nonumber \\
&~~\overset{\mathrm{(viii)}}{\ge} \frac{1}{|\Upsilon|} \sum_{\pi_{(m-k)} \in \Upsilon} \left[1 - \exp\left(- \sum_{\ell = 1}^{m-k} \Pr\left\{ \mathscr{E}_{\pi_{(m-k)}}^{(\ell)} \right\}\right)\right], \nonumber  \\
&~~\ge \frac{1}{|\Upsilon|} \sum_{\pi_{(m-k)} \in \Upsilon_{\varepsilon}} \left[1 - \exp\left(- \sum_{\ell = 1}^{m-k} \Pr\left\{ \mathscr{E}_{\pi_{(m-k)}}^{(\ell)} \right\}\right)\right],  \nonumber \\
&~~\overset{\mathrm{(ix)}}{\ge} \frac{1}{|\Upsilon|} \sum_{\pi_{(m-k)} \in \Upsilon_{\varepsilon}} \left[1 - \exp\left(-(m-k)\varepsilon \, P_{\mathrm{e}}(n, \lfloor k \varepsilon \rfloor ,p)\right)\right],  \nonumber \\
&~~> (1-2\varepsilon) \left[1 - \exp\left(-(m-k)\varepsilon \, P_{\mathrm{e}}(n, \lfloor k \varepsilon \rfloor ,p)\right)\right], \label{eq:ErrPob_AB_LB_v2}
\end{align}
where $\mathrm{(viii)}$ follows from \eqref{eq:UnionOfEvents_LB}, and $\mathrm{(ix)}$ follows using \eqref{eq:Def_UpsilonVarepsilon} as $\sum_{\ell = 1}^{m-k}\Pr\{ \mathscr{E}_{\pi_{(m-k)}}^{(\ell)}\!\} > (m-k)\varepsilon \, P_{\mathrm{e}}(n, \lfloor k \varepsilon \rfloor ,p)$ for every $\pi_{(m-k)} \in \Upsilon_{\varepsilon}$. Finally, combining \eqref{eq:Def_D_v2}, \eqref{eq:Min_D_v2}, with the fact that~\eqref{eq:ErrPob_AB_LB_v2} holds for all codebooks $\mathcal{C} \in \mathscr{C}(n,m)$, we obtain the lower bound on $\underline{D}(n,m,p,k)$ in \eqref{eq:JD_ErrProb_LB_v6_new}. 
\end{IEEEproof}

We remark that~\eqref{eq:JD_ErrProb_LB_v6_new} provides a strict improvement over the corresponding bound in~\eqref{eq:JD_ErrProb_LB_v6}. This is because the bound in \eqref{eq:JD_ErrProb_LB_v6_new} exploits {\em mutual independence} of events $\left\{\pi_{(m-k)} \to \pi_{(m-k)}^{(\ell \to \jmath)}\right\}$, while the bound in \eqref{eq:JD_ErrProb_LB_v6} only uses their {\em pairwise} independence.

\section{Continuity of the Reliability Function}
\label{App:ContinuityOfRF}
Consider a DMC with input alphabet $\mathcal{X}$, output alphabet $\mathcal{Y}$, and transition matrix $W$. Then, the (operationally defined) reliability function of the DMC is defined as~\cite{GallagerBook68}
\begin{equation}
E(R,W) \triangleq \limsup_{n \to \infty} -\frac{1}{n} \log P_{\mathrm{e}}(n,2^{nR},W) , \label{eq:Def_RF_DMC}
\end{equation}
where $P_{\mathrm{e}}(n,2^{nR},W)$ denotes the minimum error probability over all  length-$n$ block codes with $2^{nR}$ codewords. 

 We   prove that $E(R,W)$ is continuous at any  $R > R_{\infty}$, where $R_{\infty}$ is defined as the smallest $R \ge 0$ for which the sphere-packing exponent $E_{\SP}(R,W)$ is finite~\cite[p.~69]{Shannon1967_PartI}. Note that $R_{\infty} > 0$ if and only if each output $y \in \mathcal{Y}$ is unreachable from at least one input symbol in $\mathcal{X}$~\cite[p.~70]{Shannon1967_PartI}, and hence $R_{\infty} = 0$ for  BSC($p$). In the following, for brevity, we   suppress the   dependence of several quantities on $W$, e.g., we denote $E(R,W)$, $E_{\SP}(R,W)$, and $P_{\mathrm{e}}(n,2^{nR},W)$ by $E(R)$, $E_{\SP}(R)$, and $P_{\mathrm{e}}(n,2^{nR})$, respectively.
\begin{lemma}
	The reliability function $E(R)$ is continuous at any $R > R_{\infty}$.
\end{lemma}
\begin{IEEEproof}
	Let $P_{\mathrm{e}}(n, M, L)$ denote the minimum error probability for the given channel minimized over all codes with $M$ code words of length $n$ and all {\em list decoding} schemes with list size $L$. Then, \cite[Thm.~1]{Shannon1967_PartI} states that
	\begin{equation}
	P_{\mathrm{e}}(n_1 \! +\! n_2, M, L_2) \! \ge \! P_{\mathrm{e}}(n_1, M, L_1) \, P_{\mathrm{e}}(n_2, L_1 \! +\!  1, L_2). \label{eq:LB_ListDecodingErrProb}
	\end{equation}
	Note that when the list size is $L=1$, then the list-decoding error corresponds to ordinary decoding error, and for $M = 2^{nR}$, we have $P_{\mathrm{e}}(n, M, L=1) = P_{\mathrm{e}}(n, M) = P_{\mathrm{e}}(n, 2^{nR})$.

	We will employ \eqref{eq:LB_ListDecodingErrProb} for proving the continuity of $E$ at a point $R>R_\infty$. Fix $\delta \in (0,1)$ and let $n_1 = \delta n$, $n_2 = (1-\delta)n$, $L_2 = 1$  in \eqref{eq:LB_ListDecodingErrProb} to obtain
	\begin{align}
	P_{\mathrm{e}}(n,M) &\ge P_{\mathrm{e}}\left(\delta n, M, L_1\right) \, P_{\mathrm{e}}\left((1-\delta)n, L_1 + 1\right), \nonumber \\
	&\ge P_{\mathrm{e}}\left(\delta n, M, L_1\right) \, P_{\mathrm{e}}\left((1-\delta)n, L_1\right). \label{eq:LB_Pe}
	\end{align}
	Let $M = 2^{nR}$, $R' = R - R_{\infty}$, and $L_1 = 2^{n(1-\delta)(R + \delta R')}$. Then $M/L_1 = 2^{\delta n (R_{\infty} + \delta R')}$ and it follows from \cite[Thm.~2]{Shannon1967_PartI}
	\begin{equation}
	\limsup_{n \to \infty} -\frac{1}{n} \log P_{\mathrm{e}}\left(\delta n, M, L_1\right) \, \le \, \delta \, E_{\SP}(R_{\infty} + \delta R'). \label{eq:LD_Bound}
	\end{equation}
	As $L_1 = 2^{(1-\delta)n(R + \delta R')}$, it follows from \eqref{eq:Def_RF_DMC} that
	\begin{equation}
	\limsup_{n \to \infty} - \frac{1}{n} \log P_{\mathrm{e}}\left((1-\delta)n, L_1\right) \, = \, (1-\delta) \, E\left(R + \delta R'\right). \label{eq:AsympEq1}
	\end{equation}
	The $\limsup$ operator is subadditive, i.e., for two sequences $(a_n)$ and $(b_n)$, we have $\limsup_{n \to \infty}(a_n + b_n) \le \limsup_{n \to \infty}a_n + \limsup_{n \to \infty}b_n$. Combining this fact with~\eqref{eq:LB_Pe}, we get
	\begin{align*}
	\limsup_{n\to\infty}-\frac{1}{n}\log P_{\mathrm{e}}(n,M) &\le \limsup_{n\to\infty}-\frac{1}{n}\log P_{\mathrm{e}}\left(\delta n, M, L_1\right) \\
	&+ \limsup_{n\to\infty}-\frac{1}{n}\log P_{\mathrm{e}}\left((1-\delta)n, L_1\right).
	\end{align*}
	Applying \eqref{eq:LD_Bound} and~\eqref{eq:AsympEq1} to the above inequality, we get
	\begin{equation}
	E(R) \, \le \, \delta\, E_{\SP} (R_{\infty} + \delta R') + (1-\delta)\, E\left(R +\delta R'\right) . \label{eq:UB_RF_v1}
	\end{equation}
	As \eqref{eq:UB_RF_v1} holds for all $\delta > 0$, we have
	\begin{align}
	E(R) &\le \liminf_{\delta \downarrow 0} \left[\delta\, E_{\SP} (R_{\infty} + \delta R') + (1-\delta)\, E\left(R +\delta R'\right)\right], \nonumber \\
	&\overset{\mathrm{(i)}}{=}  \liminf_{\delta \downarrow 0} \left[(1-\delta)\, E\left(R +\delta R'\right)\right], \nonumber \\
	&= \liminf_{\delta \downarrow 0} E\left(R +\delta R'\right), \label{eq:RF_Ineq1} 
	\end{align}
	where $\mathrm{(i)}$ follows because $\lim_{\delta \downarrow 0} \left[\delta\, E_{\SP} (R_{\infty} + \delta R')\right] = 0$  as $E_{\SP}(R_\infty)$ is finite. Next, as $E(R)$ is non-increasing in $R$, 
	\begin{equation}
	E(R +\delta R') \le E(R), \label{eq:RF_Ineq2} 
	\end{equation}
	and it follows from \eqref{eq:RF_Ineq2} that
	\begin{equation}
	\limsup_{\delta \downarrow 0} E(R +\delta R') \le E(R). \label{eq:RF_Ineq3}
	\end{equation}
	Now \eqref{eq:RF_Ineq1} and \eqref{eq:RF_Ineq3} imply that $\lim_{\delta \downarrow 0} E(R +\delta R')$ exists, and
	\begin{equation}
	E(R) = \lim_{\delta \downarrow 0} E(R +\delta R') = \lim_{\delta \downarrow 0} E(R + \delta). \label{eq:RF_Ineq4}
	\end{equation}
	The above argument  leading to~\eqref{eq:RF_Ineq4} shows that $E(R)$ {\em is right continuous}. To prove {\em left continuity}, we again fix $R > R_{\infty}$ and choose $\delta \in (0,1)$. Let $R' = R - R_{\infty}$, $M = 2^{n\left(R_{\infty} + [R'/(1+\delta)]\right)}$ and $L_1 = 2^{n(1-\delta)(R_{\infty} + R')}$. Then, $M/L_1 = 2^{ (\delta n)\left(R_{\infty} + R'\delta/(1+\delta)\right)}$. From \cite[Thm.~2]{Shannon1967_PartI}, we have
	\begin{equation}
	\limsup_{n \to \infty} -\frac{1}{n} \log P_{\mathrm{e}}\left(\delta n, M, L_1\right) \le \delta E_{\SP}(R_{\infty} + R' \delta/(1+\delta)). \label{eq:LD_Bound_v2}
	\end{equation}
	As $L_1 = 2^{(1-\delta)n R}$, it follows from \eqref{eq:Def_RF_DMC} that
	\begin{equation}
	\limsup_{n \to \infty} - \frac{1}{n} \log P_{\mathrm{e}}\left((1-\delta)n, L_1\right) \, = \, (1-\delta) \, E(R). \label{eq:AsympEq2}
	\end{equation}
	Combining \eqref{eq:LB_Pe}, \eqref{eq:LD_Bound_v2}, and \eqref{eq:AsympEq2}, we obtain
	\begin{equation*}
	E\left(R_{\infty} + \frac{R'}{1+\delta}\right) \le \delta E_{\SP}\left(R_{\infty} + \frac{R' \delta}{1+\delta}\right) + (1-\delta)\, E(R) . 
	\end{equation*}
	The above inequality can be equivalently expressed as
	{\small
	\begin{equation*}
	E(R) \ge \frac{1}{1-\delta}\, E\left(R_{\infty} + \frac{R'}{1+\delta}\right) - \frac{\delta}{1-\delta}\, E_{\SP}\left(R_{\infty} + \frac{R'\delta}{1+\delta}\right).
	\end{equation*}
	}
	\normalsize
	The above relation holds for all $\delta > 0$, and hence
	\begin{align*}
	E(R) \ge \limsup_{\delta \downarrow 0} &\Bigg(\frac{1}{1-\delta}\, E\left(R_{\infty} + \frac{R'}{1+\delta}\right) \\*
	&- \frac{\delta}{1-\delta}\, E_{\SP}\left(R_{\infty} + \frac{R'\delta}{1+\delta}\right) \Bigg).
	\end{align*}
	Now, $\lim_{\delta \downarrow 0} \left(\delta/(1-\delta)\right) E_{\SP} \left(R_{\infty} + R'\delta/(1+\delta)\right) = 0$ as $E_{\SP}(R_{\infty})$ is finite, and it follows from above inequality that 
	\begin{align}
	E(R) &\ge \limsup_{\delta \downarrow 0}\left(\frac{1}{1-\delta}\, E\left(R_{\infty} + \frac{R'}{1+\delta}\right)\right), \nonumber \\
	&= \limsup_{\delta \downarrow 0} E\left(R_{\infty} + \frac{R'}{1+\delta}\right), \nonumber \\
	&= \limsup_{\delta \downarrow 0} E\left(R_{\infty} + R'(1-\delta)\right), \nonumber \\
	&= \limsup_{\delta \downarrow 0} E\left(R - \delta\right). \label{eq:RF_Ineq5} 
	\end{align}
	Further, as $E(R)$ is non-increasing in $R$, we have 
	\begin{equation}
	\liminf_{\delta \downarrow 0} E\left(R - \delta\right) \ge E(R). \label{eq:RF_Ineq7}
	\end{equation}
	Combining \eqref{eq:RF_Ineq5} and \eqref{eq:RF_Ineq7}, we get
	\begin{equation}
	E(R) = \lim_{\delta \downarrow 0} E(R - \delta), \label{eq:RF_Ineq8}
	\end{equation}
	which proves that $E(R)$ is left continuous, and the proof is complete by combining \eqref{eq:RF_Ineq4} and \eqref{eq:RF_Ineq8}.
\end{IEEEproof}

\section{Proof of Lemma~\ref{Lem:ConvergenceToRF}} \label{App:ConvergenceToRF}
\begin{IEEEproof}
	Choose $R>0$ and $\epsilon > 0$. As $E(\tilde{R},p)$ is continuous in $\tilde{R}$ (see Appendix~\ref{App:ContinuityOfRF}), there exists $0 < \delta < R$ such that $|E(\tilde{R},p) - E(R,p)| < \epsilon$ for all $|\tilde{R} - R| \le  \delta$. Now, as $R_n$ converges to $R$, there exists an $N$ such that $|R_n - R| \le \delta$ for all $n \ge N$. Furthermore, as $E(n,R,p)$ is non-increasing in $R$,
	\begin{equation*}
	E(n,R_n,p) \le E(n,R-\delta,p), \mathrm{~~for~} n\ge N.
	\end{equation*}
	From the above inequality, it follows that
	\begin{align}
	\limsup_{n \to \infty} E(n,R_n,p) &\le \limsup_{n \to \infty} E(n,R-\delta,p), \nonumber \\
	&= E(R-\delta,p), \nonumber \\
	&< E(R,p) + \epsilon   \label{eq:E_Rn_UB} 
	%&< E(R,p) + \epsilon.
	\end{align}
	As $E(n,R,p)$ is non-increasing in $R$, we have
	\begin{equation*}
	E(n,R_n,p) \ge E(n,R + \delta,p), \mathrm{~~for~} n\ge N.
	\end{equation*}
	From the above inequality, it follows that
	\begin{align}
	\limsup_{n \to \infty} E(n,R_n,p) &\ge \limsup_{n \to \infty} E(n,R+\delta,p), \nonumber \\
	&= E(R+\delta,p), \nonumber \\
	&\ge E(R,p) - \epsilon . \label{eq:E_Rn_LB}
	%&> E(R,p) - \epsilon. 
	\end{align}
	The proof is complete by observing that \eqref{eq:E_Rn_UB} and \eqref{eq:E_Rn_LB} hold for all $\epsilon>0$.
\end{IEEEproof}

\section{Proof of Theorem~\ref{Thm:MainResult}} \label{App:MainResult}
\begin{IEEEproof}
We first show that $E_{\underline{D}}(R,p,\alpha) \ge |E(R,p) - R|^+$. Towards this, we note that when $m = 2^{nR}$ and $k = \alpha m$, for a given $0 < \alpha < 1$, then we have
\begin{align}
\lim_{n \to \infty}	-\frac{1}{n} \log (m-k) &= \left(\lim_{n \to \infty} -\frac{1}{n} \log (1-\alpha)\right) - R, \nonumber \\
&= -R. \label{eq:m_minus_k_asym_v1}
\end{align} 
Combining \eqref{eq:ID_UB_ErrProb_v2}, \eqref{eq:ErrExp_Dmin}, and \eqref{eq:m_minus_k_asym_v1}, we get
\begin{align}
E_{\underline{D}}(R,p,\alpha) &\ge \left| \left(\limsup_{n \to \infty} -\frac{1}{n} \log P_{\mathrm{e}}(n,m,p) \right) - R \,\right|^+ , \nonumber \\
&= \left|E(R,p) - R \right|^+, \label{eq:BeeErrExpAbsentee_LB_v1}
\end{align}
where the last equality follows from \eqref{eq:ErrExp_Pe_n} and \eqref{eq:ErrExp_Pe}.

Next, we show that $E_{\underline{D}}(R,p,\alpha) \le |E(R,p) - R|^+$ by applying Lemma~\ref{Lem:ErrProb_LB_JD}. Choose $\varepsilon = 1/4$, and define
\begin{align}
\hat{R}_n \triangleq \frac{1}{n} \log (\lfloor k \varepsilon \rfloor ) . \label{eq:R_hat_n_Def}
\end{align}
For $k > 8$, we have $k \varepsilon/2 < \lfloor k \varepsilon \rfloor  \le k \varepsilon$. Thus, when $k = \alpha m$ and $m = 2^{nR}$, we get
\begin{align}
R + \frac{1}{n} \log \left(\frac{\alpha \varepsilon}{2}\right) &< \hat{R}_n \le R + \frac{1}{n} \log (\alpha \varepsilon) ,
\end{align}
which implies that 
\begin{align}
R &= \lim_{n \to \infty} \hat{R}_n . \label{eq:lim_R_hat_n}
\end{align}
Combining the facts that $\lim_{n \to \infty}\tfrac{1}{n}\log\left((1-2\varepsilon)/2\right) = 0$, $\lim_{n \to \infty}\tfrac{1}{n} \log\left((m-k) \varepsilon\right) = R$, with \eqref{eq:ErrExp_Dmin}, \eqref{eq:JD_ErrProb_LB_v6}, \eqref{eq:ErrExp_Pe_n}, and~\eqref{eq:R_hat_n_Def}, we get
\begin{align}
E_{\underline{D}}(R,p,\alpha) &\le \left|\limsup_{n \to \infty} E(n,\hat{R}_n, p) - R \right|^+ , \nonumber \\
&= \left|E(R,p) - R \right|^+, \label{eq:BeeErrExpAbsentee_UB_v1}
\end{align}
where the last equality follows from \eqref{eq:lim_R_hat_n} and \eqref{eq:ErrExp_Pe}. The proof is now complete by combining \eqref{eq:BeeErrExpAbsentee_LB_v1} and \eqref{eq:BeeErrExpAbsentee_UB_v1}.
\end{IEEEproof}

\section{Proof of Theorem~\ref{Thm:CapBeeProblem}} \label{app:CapBeeProblem}
\begin{IEEEproof}
	We first prove the {\em direct} part $C_{\underline{D}}(p,\alpha,\epsilon) \ge R_p^*$.
	If $R < R_p^*$, then it follows from \eqref{eq:ErrExp_Dmin_Exact} and the definition of $R_p^*$ that $E_{\underline{D}}(R,p,\alpha)$ is \emph{strictly} positive. From \eqref{eq:ErrExp_Dmin} it follows that there exist infinitely many $n$ for which 
	\begin{equation*}
	-\frac{1}{n} \log \underline{D}(n,2^{nR},p,\alpha 2^{nR}) > E_{\underline{D}}(R,p,\alpha)/2.
	\end{equation*}
	In other words, when $R<R_p^*$, $\underline{D}(n,2^{nR},p,\alpha 2^{nR}) < 2^{-n E_{\underline{D}}(R,p,\alpha) / 2}$. Thus when $R < R_p^*$, we have $$\liminf_{n \to \infty} \underline{D}(n,2^{nR},p,\alpha 2^{nR}) = 0.$$ Therefore, any rate less than $R_p^*$ is achievable and it follows from the definition of capacity in~\eqref{eq:Def_CapBeeProblem} that $C_{\underline{D}}(p,\alpha,\epsilon) \ge R_p^*$.
	
	Next, we will apply the bound in \eqref{eq:JD_ErrProb_LB_v6_new} to prove the converse part $C_{\underline{D}}(p,\alpha,\epsilon) \le R_p^*$. This is a  {\em strong converse} statement, i.e., for rates $R > R_p^*$, the error probability $\underline{D}(n,2^{nR},p,\alpha 2^{nR})$ tends to 1 as $n \to \infty$. Consider a rate $R$ that satisfies $R > R_p^*$, and define $\Delta_R \triangleq R - E(R,p)$. Then it follows from the definition of $R_p^*$, and the fact $E(R,p)$ is non-increasing in $R$, that $\Delta_R > 0$. Define $\varepsilon_n \triangleq \tfrac{1}{n}$, and let $n$ be sufficiently large such that $k = \alpha 2^{nR} > 2n = 2/\varepsilon_n$. Now define $\hat{R}_n$ to be
	\begin{equation}
	\hat{R}_n \triangleq \frac{1}{n} \log (\lfloor k \varepsilon_n \rfloor ) .\label{eq:Def_R_hat_n_v2}
	\end{equation}
	Then, we have
	\begin{align}
	R + \frac{1}{n} \log \left(\frac{\alpha}{2n}\right) &< \hat{R}_n \le R + \frac{1}{n} \log \left(\frac{\alpha}{n}\right) ,\nonumber \\
	R &= \lim_{n \to \infty} \hat{R}_n . \label{eq:lim_R_hat_n_v2}
	\end{align}
	It follows from \eqref{eq:lim_R_hat_n_v2} and \eqref{eq:ErrExp_Pe} that
	\begin{equation*}
	\limsup_{n \to \infty} E(n,\hat{R}_n, p) = E(R,p).
	\end{equation*}
	As $\Delta_R > 0$, the above equation implies that there exists an $N$ such that for all $n \ge N$, we have
	\begin{equation}
	E(n,\hat{R}_n, p) < E(R,p) + \frac{\Delta_R}{2} . \label{eq:ErrExpFiniteLength_UB}
	\end{equation}
	Combining \eqref{eq:ErrExp_Pe_n}, \eqref{eq:Def_R_hat_n_v2}, and \eqref{eq:ErrExpFiniteLength_UB}, for $n \ge N$, we obtain
	\begin{equation}
	P_{\mathrm{e}}(n,\lfloor k \varepsilon_n \rfloor,p) > 2^{-n \left(E(R,p) + \left(\Delta_R/2\right)\right)}. \label{eq:Pe_n_LB_v2}
	\end{equation}
	Now, if we define $\beta_n$ as
	\[
	\beta_n \triangleq -\frac{1}{n} \log \left((1-\alpha) \varepsilon_n \right) ,
	\]
 	then we have $\beta_n > 0$, while $\lim_{n \to \infty} \beta_n = 0$. Thus, we have
	\begin{equation}
	(m-k)\varepsilon_n = m(1-\alpha)\varepsilon_n = 2^{n \left(R - \beta_n\right)}. \label{eq:m_minus_k_new}
	\end{equation}
	Combining \eqref{eq:Pe_n_LB_v2} and \eqref{eq:m_minus_k_new}, for all $n \ge N$, we have 
	\begin{align}
	(m-k)\varepsilon_n\, P_{\mathrm{e}}(n,\lfloor k \varepsilon_n \rfloor,p) &> 2^{n \left(R - E(R,p) - \left(\Delta_R/2\right) - \beta_n\right)}, \nonumber \\
	&= 2^{n \left( (\Delta_R/2) - \beta_n\right)}. \label{eq:ConverseProof_Ineq1_v2}
	\end{align}
	As \eqref{eq:JD_ErrProb_LB_v6_new} holds for all $0 < \varepsilon < 1/2$ and $k > 1/\varepsilon$, replacing $\varepsilon$ with $\varepsilon_n = \tfrac{1}{n}$ in \eqref{eq:JD_ErrProb_LB_v6_new}, we get for $n > N$ that
	\begin{align}
	&\underline{D}(n,2^{nR},p,\alpha 2^{nR}) \nonumber \\
	&~~> (1-2\varepsilon_n) \left[1 - \exp\left(-(m-k)\varepsilon_n \, P_{\mathrm{e}}(n, \lfloor k \varepsilon_n \rfloor ,p)\right)\right], \nonumber \\
	&~~> (1-2\varepsilon_n) \big[1 - \exp\big(-2^{n \left( (\Delta_R/2) - \beta_n\right)}\big)\big], \label{eq:JD_ErrProb_LB_v7_new}
	\end{align}
	where \eqref{eq:JD_ErrProb_LB_v7_new} follows from \eqref{eq:ConverseProof_Ineq1_v2}. Now, as $\beta_n = o(1)$, there exists $\hat{N}$ such that for all $n \ge \hat{N}$, we have $\Delta_R/2 - \beta_n > 0$. Further, as $\beta_n$ is a decreasing function of $n$, it follows that
	\begin{equation}
	\lim_{n \to \infty} \big[1 - \exp\big(-2^{n \left( (\Delta_R/2) - \beta_n\right)}\big)\big] = 1. \label{eq:ConverseProof_Ineq2_v2}
	\end{equation}
	As $\lim_{n \to \infty} (1-2\varepsilon_n) = 1$, combining \eqref{eq:JD_ErrProb_LB_v7_new} and \eqref{eq:ConverseProof_Ineq2_v2} with the fact that $\underline{D}(n,2^{nR},p,\alpha 2^{nR})$ is upper bounded by 1, we obtain the following important result 
	\begin{equation}
	\lim_{n \to \infty} \underline{D}(n,2^{nR},p,\alpha 2^{nR}) = 1, \;\mbox{for}\; R > R_p^*, \label{eq:ConverseProof_Ineq3_v2}
	\end{equation}
	thereby showing that $C_{\underline{D}}(p,\alpha,\epsilon) \le R_p^*$, and completing the proof of the strong converse.	
\end{IEEEproof}

\section{Proof of Theorem~\ref{Thm:ErrExp_AlphaToZero}} \label{App:ErrExp_AlphaToZero}
\begin{figure}[t]
	\centering
	\includegraphics[width=0.48\textwidth]{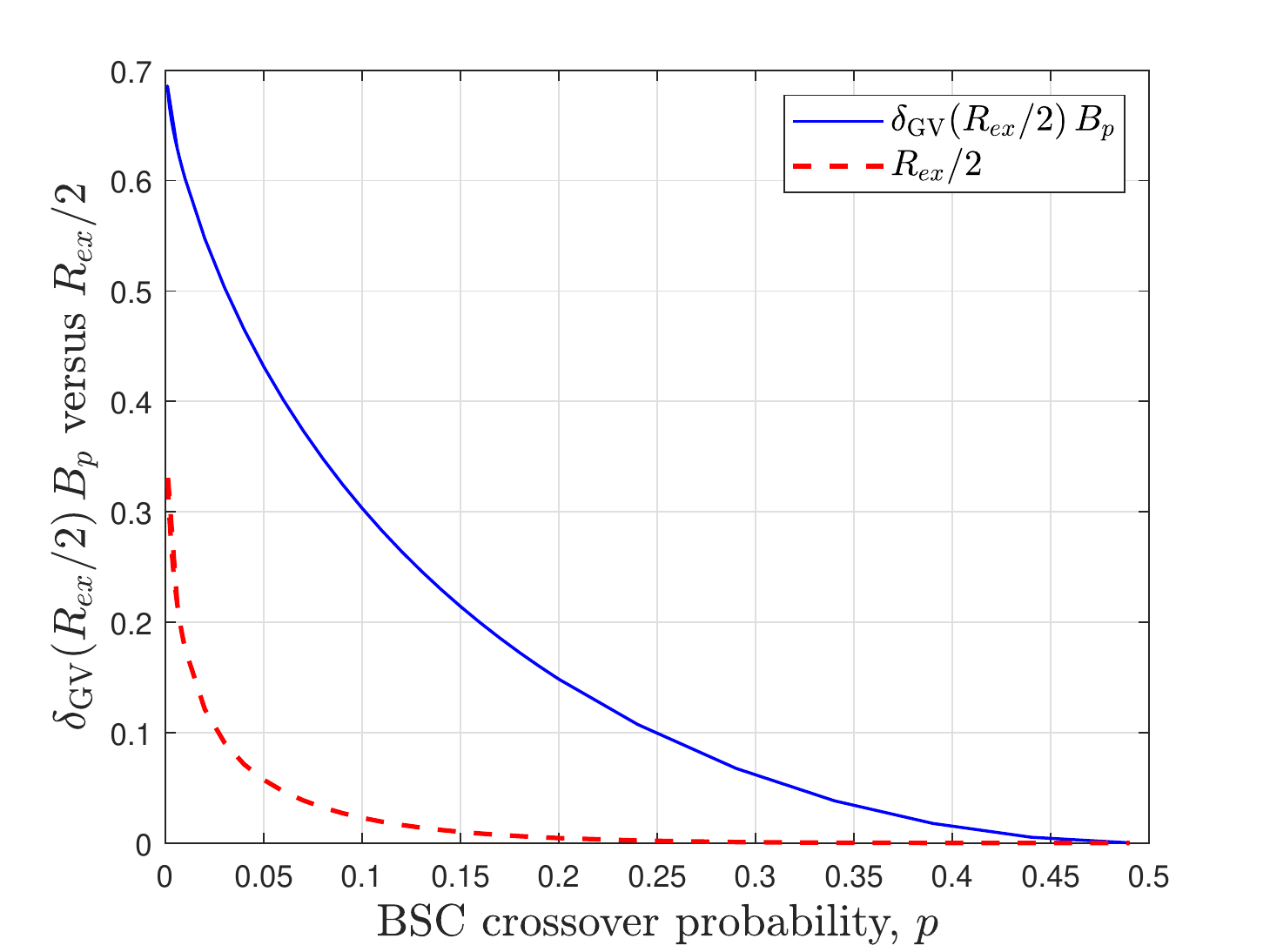}
	\caption{Plot demonstrating $\delta_{\GV}(R_{\mathrm{ex}}/2) B_p \, > \, R_{\mathrm{ex}}/2$.}
	\label{Fig:ErrExp_at_HalfRex}
\end{figure}

\begin{figure}[t]
	\centering
	\includegraphics[width=0.48\textwidth]{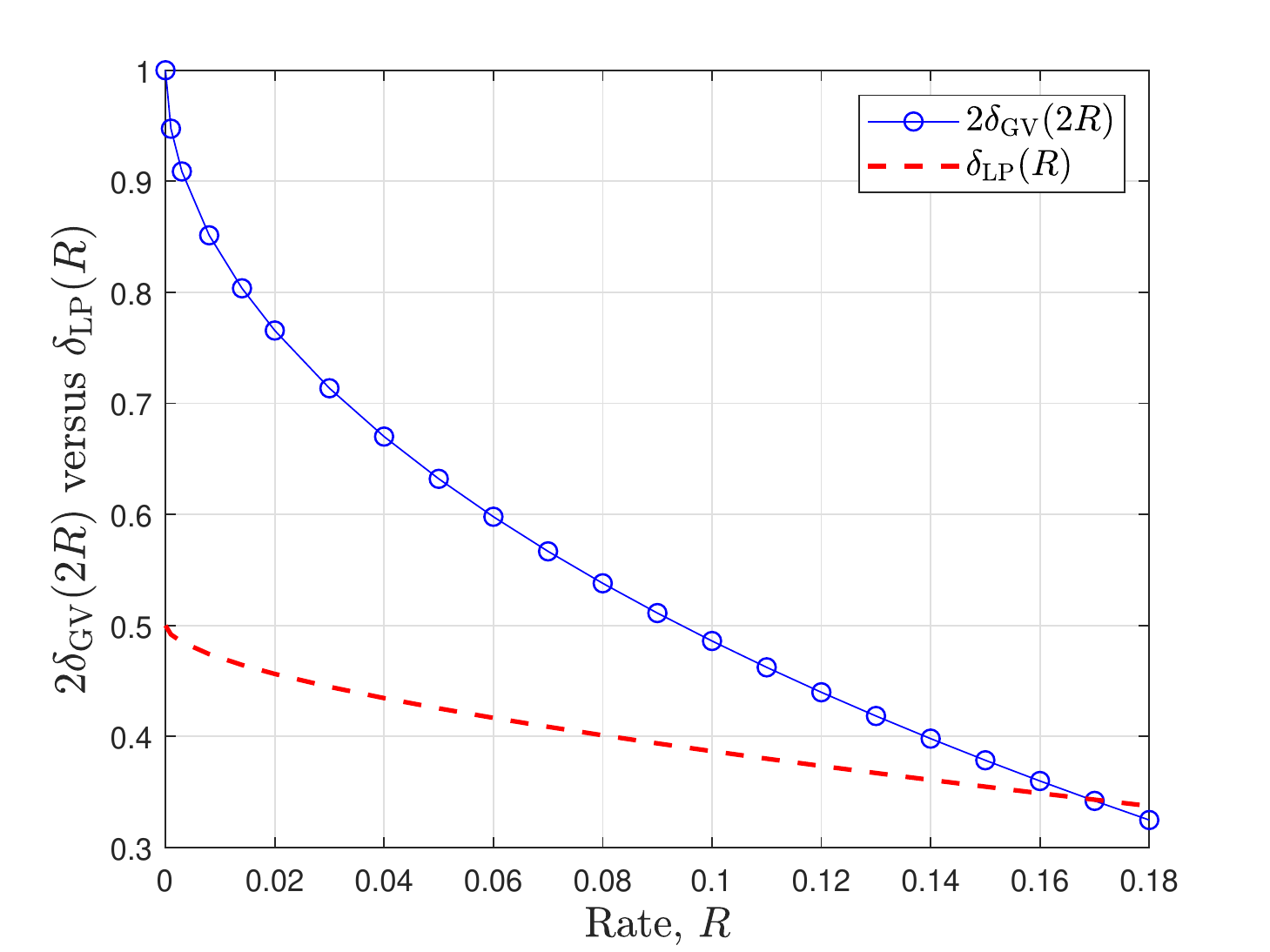}
	\caption{$\delta_{\LP}(R) < 2\delta_{\GV}(2R)$ for $0 < R < 0.169$.}
	\label{Fig:deltaGV_deltaLP}
\end{figure}

\begin{IEEEproof}
We first show that $E(R,p) - R > 0$ when $0 < R < R_{\mathrm{ex}}/2$. Towards this, we note from \eqref{eq:BoundReliabilityFunction} and \eqref{eq:ETLC_Def} that $E(R_{\mathrm{ex}}/2,p) \ge \delta_{\GV}(R_{\mathrm{ex}}/2) B_p$. Now, it can be numerically verified that $\delta_{\GV}(R_{\mathrm{ex}}/2) B_p > \left(R_{\mathrm{ex}}/2\right)$ when $0 < p < 0.5$ (see Fig.~\ref{Fig:ErrExp_at_HalfRex}), and hence $E(R_{\mathrm{ex}}/2,p) > R_{\mathrm{ex}}/2$. As $E(R,p)$ is non-increasing in $R$, it follows that
\begin{equation}
E(R,p) - R > 0, \;\;\mbox{when}\;\; 0 < R < R_{\mathrm{ex}}/2. \label{eq:Erp_at_HalfRex}
\end{equation}
Now, for $0 < R < \min\left\{0.169,\, R_{\mathrm{ex}}/2\right\}$, we have
\begin{align*}
\lim_{\alpha \downarrow 0} E_{\underline{D}}(R,p,\alpha) &\overset{\mathrm{(i)}}{=} |E(R,p) - R|^+ ,\\
&\overset{\mathrm{(ii)}}{=} E(R,p) - R , \\
&\overset{\mathrm{(iii)}}{\le}  \delta_{\LP}(R) B_p - R , \\
&\overset{\mathrm{(iv)}}{<} 2 \delta_{\GV}(2R) B_p - R, \\
&< 2 \delta_{\GV}(2R) B_p , \\
&\overset{\mathrm{(v)}}{\le} E_{\underline{D}}(R,p) ,
\end{align*}
where $\mathrm{(i)}$ follows from Thm.~\ref{Thm:MainResult} and the fact that $E_{\underline{D}}(R,p,\alpha)$ is constant for $0 < \alpha < 1$, \ $\mathrm{(ii)}$ follows from \eqref{eq:Erp_at_HalfRex}, \ $\mathrm{(iii)}$ follows from \eqref{eq:RF_UB2}, \ $\mathrm{(iv)}$ follows from the fact that $\delta_{\LP}(R) < 2\delta_{\GV}(2R)$ for $0 < R < 0.169$ (see Fig.~\ref{Fig:deltaGV_deltaLP}),\ and $\mathrm{(v)}$ follows from \cite[Theorem~4]{Tandon19_Bee_TCOM}.\end{IEEEproof}

\section*{Acknowledgement}
The authors acknowledge discussions with Po-Ning Chen, Fady Alajaji, and Mariam Harutyunyan, on the continuity of the reliability function for discrete memoryless channels.


\end{document}